\documentclass{aa}  
\usepackage{graphicx}
\usepackage{txfonts}
\usepackage{amsmath} 
\usepackage{amssymb} 
\usepackage{xcolor}
\usepackage[]{geometry}
\usepackage{txfonts}

\usepackage[]{hyperref}

\usepackage{verbatim}
\usepackage{enumitem}
\usepackage{epsfig}
\usepackage{multirow}
\usepackage{tablefootnote}
\usepackage{stfloats}
\usepackage{makecell}
\usepackage{graphicx}

\usepackage{booktabs}
\usepackage{rotating}
\usepackage{subcaption}
\makeatletter
\renewcommand{\fnum@table}{\textbf{Table \thetable}}
\renewcommand{\fnum@figure}{\textbf{Figure \thefigure}}
\makeatother
\usepackage{ulem}
\usepackage{setspace}
\usepackage{placeins}

\def\erosita{{\it eROSITA}}
\def\xmm{{\it XMM-Newton}}
\def\chandra{{\it Chandra}}
\def\rosat{{\it ROSAT}}
\def\gaia{{\it Gaia}}

\begin{document}

   \title{The average X-ray spectrum of the volume-complete M and F, G, K type stars sample within 10 pc of the Sun}

    \author{
    Xueying Zheng\inst{1}\thanks{emails to: zhengxy@mpe.mpg.de},
    Gabriele Ponti\inst{2,1,3},
    Nicola Locatelli \inst{2},
    Beate Stelzer\inst{4}, 
    Enza Magaudda\inst{4},
    Konrad Dennerl\inst{1},
    Michael Freyberg\inst{1},
    Jeremy Sanders\inst{1},
    Marilena Caramazza\inst{4}, 
    Manami Sasaki\inst{5},
    Andrea Merloni\inst{1},
    Jan Robrade \inst{6},
    Teng Liu\inst{1,7,8},
    He-shou Zhang\inst{2},
    Martin G. F. Mayer\inst{5},
    Yi Zhang \inst{1},
    Michael C. H. Yeung \inst{1},
    Werner Becker\inst{1}
    }

\institute{
     Max-Planck-Institut f{\"u}r extraterrestrische Physik, Gießenbachstraße 1, 85748 Garching bei M\"unchen, Germany
\and INAF-Osservatorio Astronomico di Brera, Via E. Bianchi 46, I-23807 Merate (LC), Italy
\and Como Lake centre for AstroPhysics (CLAP), DiSAT, Università dell’Insubria, via Valleggio 11, 22100 Como, Italy
\and Institut für Astronomie und Astrophysik, Eberhard Karls Universität Tübingen, Sand 1, 72076 Tübingen, Germany
\and Dr. Karl Remeis Observatory, Erlangen Centre for Astroparticle Physics, Friedrich-Alexander-Universit\"at Erlangen-N\"urnberg Sternwartstraße 7, 96049 Bamberg, Germany
\and Hamburger Sternwarte, Gojenbergsweg 112, 21029 Hamburg, Germany
\and Department of Astronomy, University of Science and Technology of China, Hefei 230026, China
\and School of Astronomy and Space Science, University of Science and Technology of China, Hefei 230026, China}
   \date{}

  \abstract
   {F, G, K and M type stars are the most abundant stellar population in the Milky Way and are expected to contribute to its diffuse X-ray emission. Yet their intrinsic average X-ray spectrum remains poorly constrained due to their faint X-ray luminosities, leaving their collective role in the X-ray background of the Milky Way uncertain.}   
   {We aim to derive a distance-normalized average X-ray spectrum for nearby M dwarfs and FGK stars, and to characterize their ensemble spectral properties. }     
  {We analysed the volume-complete sample of M dwarfs (M0–M6) and FGK stars within 10 pc of the Sun using data from \erosita\ all-sky survey aboard the Spectrum-Roentgen-Gamma (SRG) mission (eRASS:4). Individual spectra were normalized by exposure and distance and then stacked to produce representative averages. From these spectra we derived characteristic X-ray luminosities, that are helpful to estimate the contribution of late-type stars to the soft diffuse X-ray emission of the Galaxy.}        
  {The distance-normalized emission measures yield an average X-ray luminosity of $(2.6 \pm0.1)\times 10^{27}$ erg/s for M-type stars, and $(15\pm3)\times 10^{27}$ erg/s for F, G and K-type stars in 0.2--2.0 keV. The 10 pc average spectra could be well described by a sum of three and two thermal models. The fitted temperatures and abundances remain consistent across M-star subgroups, while early-M stars are surprisingly on average less luminous than mid/late-M types. These results offer new insights into the collective X-ray properties of nearby stars in the volume-complete context, and provide motivation to further explore the link with the unresolved soft X-ray background of the Galaxy.}
   {}


\titlerunning{The average X-ray spectrum of stars in the solar neighbourhood}
\authorrunning{X.Y. Zheng  et. al}
\maketitle

%

\section{Introduction}

Late-type main-sequence stars, such as F-, G-, K-, and M-type stars, are the most numerous stellar type in the Milky Way. M dwarfs alone account for about 70\% of the entire stellar population, yet contribute only about 35\% of the total stellar mass. FGK-type stars, 
though less common ($\sim 23\%$), provide a comparable fraction of stellar mass ($\sim$35\%) \citep{Reid1997AJ, Bochanski2010AJ, Golovin2023AA, Kirkpatrick2024ApJS}.
Their X-ray emission arises from magnetically heated coronae, where plasma is confined and energized by magnetic fields generated through convection and rotation \citep{Parker1993ApJ, Gudel2004AA}.
Understanding this emission is crucial, as it serves as a proxy for their dynamo and shapes the environments of surrounding planets. 
Individually, these stars appear as faint X-ray sources even at close distances, but their sheer numbers suggest that they contribute significantly to the soft X-ray background, including the local hot bubble and Galactic Halo (\citealt{Yeung2024AA}; Ponti et al. in prep.).

A systematic census of the X-ray emission from nearby late-type stars was first carried out with the \rosat\ All-Sky Survey, led by \cite{Schmitt2004AA} who compiled the ROSAT/NEXXUS catalogue of all stars in the solar neighbourhood. Deeper observations involving \xmm, and \chandra, combined with \rosat\, have since enabled detailed spectral investigations of individual stars.
Multi-temperature collisionally ionized thermal model has provided good fits to these stellar coronae, such as in AD Leonis, EV Lac and EQ Pegasi \citep{Raymond1977ApJS, Mewe1991AARv, Robrade2005AA}.

More recent efforts have been directed towards fully characterizing the X-ray properties of volume-limited samples through systematic surveys with the help of \gaia.
Based on the \gaia\ 10-pc catalogue 
\citep{Reyle2021AA}, \citet{Caramazza2023AA} conducted the census of X-ray observations of M dwarfs within the closest 10 pc using \rosat, \xmm, and \erosita, probing the faintest levels of coronal emission.
\citet{Zhu2025AA} and \citet{Locatelli2025AA} also combined the archival X-ray data to study the distribution of luminosity of nearby stars of F to M type in different volumes. 
Together, these studies provide important constraints on the prevalence, distribution of stellar X-ray activity, with emphasis on individual luminosity measurements and their relation to other stellar properties.

In this paper, we build on these efforts by deriving a distance-normalized stacked X-ray spectrum from the complete, volume-limited sample of nearby late-type stars using data from the four sky surveys of \erosita\ (eRASS:4; see Section~\ref{sec:data}).
Our stellar sample is based on the \gaia\ 10-pc catalogue \citep{Reyle2021AA}, with the M-dwarf subsample (M0–M4) and their coordinates provided by \citet{Caramazza2023AA}, and the complementary M4–M6 and FGK-dwarf lists supplied by B. Stelzer (priv. comm.).
We extract spectra from the merged \erosita\ eRASS:4 event-files and stack them to yield an emissivity-weighted average luminosity for M stars and FGK stars separately.

The paper is structured as follows. 
The data selection procedure is described in Section~\ref{sec:sample};
Section~\ref{sec:data} presents the data processing; 
Section~\ref{sec:method} explains the stacking approach; 
Section~\ref{sec:model_fitting} introduces model fitting and spectral analysis for the averaged spectrum of M0--M6 stars within 10 pc; 
Section~\ref{sec:fittingFGK} discusses the spectral analysis of FGK-type stars within 10 pc
and finally, Section~\ref{sec:conclu} summarizes the results.

\begin{figure*}[]
    \centering
    \includegraphics[trim={0.1cm 0.0cm 0.0cm 0.1cm}, clip, width=\linewidth]{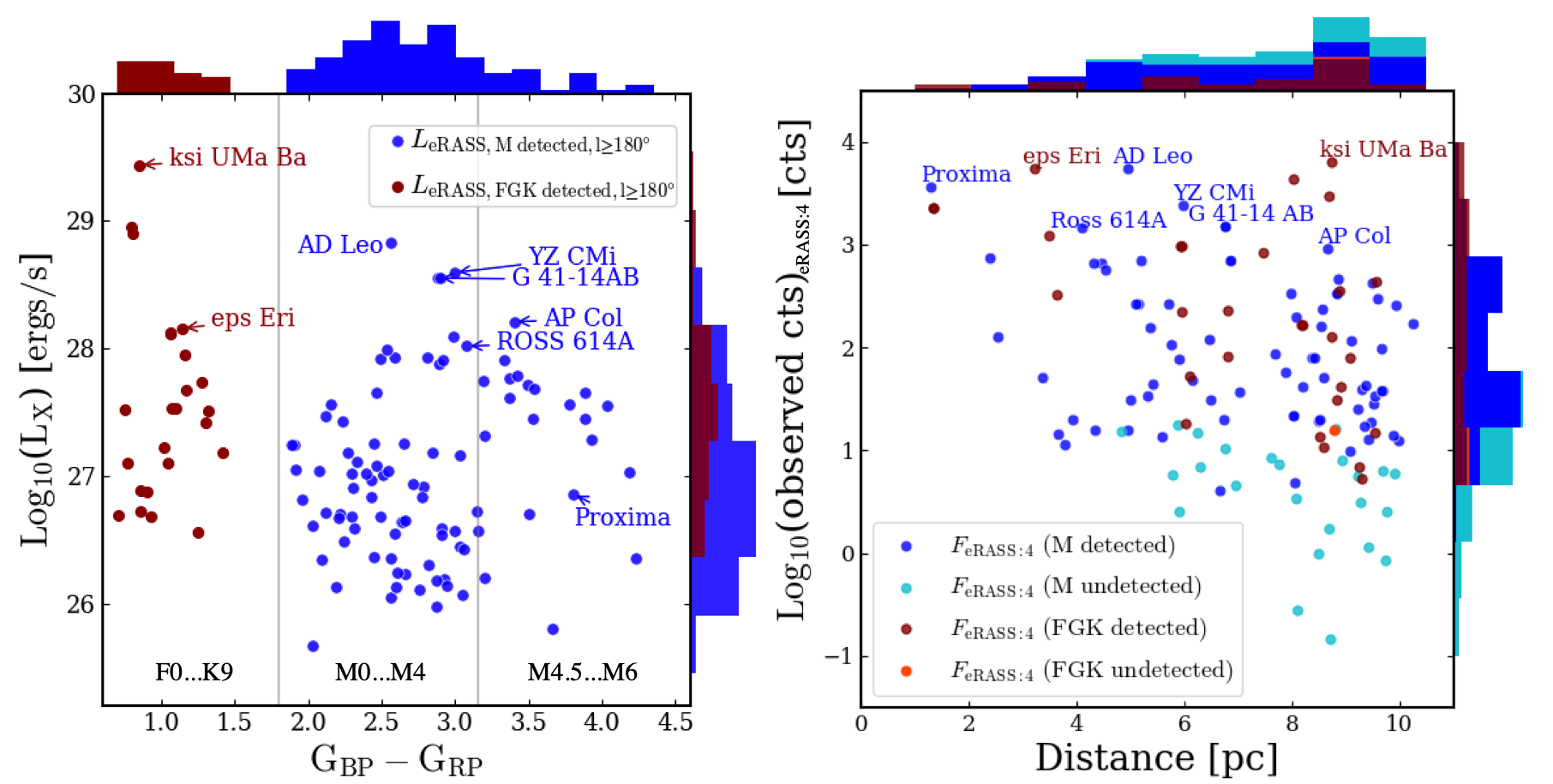}
    \caption{ Left: X-ray Luminosity (0.1--2.4 keV) versus \gaia\ $G_{\mathrm{BP}}$-$G_{\mathrm{RP}}$ colour for the eRASS1-detected M and FGK-type star \citep{Merloni2024AA} in the western Galactic hemisphere of 10pc-\gaia\ sample. 
    Filled blue circles: eRASS1 detected M~dwarfs ($l \geq 180^\circ$);
    filled red circles: eRASS1 detected FGK~stars ($l \geq 180^\circ$).
    Notable sources are labeled.  
    Right: eRASS:4 total observed counts in 0.2--2.0 keV versus the distance from the Sun for the 10pc-\gaia\ sample in western Galactic hemisphere,
    with detections and non-detections for M~dwarfs (blue, cyan) and FGK stars (brown, coral).}
    \label{fig:overview}
\end{figure*}

\section{Sample selection}\label{sec:sample}

As mentioned above, we use the volume-complete \gaia{} based catalogue of stars within $10$\,pc of the Sun \citep{Reyle2021AA} as master sample. 
From this, we adopt the F, G, K, and M0--M6 spectral type (SpT) subsample defined by Stelzer et al.(in prep.), who, following \cite{Caramazza2023AA}, assign SpT based on \gaia\ eDR3 $G_{BP} - G_{RP}$ colours using the conversion table compiled by E. Mamajek \footnote{\url{http://www.pas.rochester.edu/~emamajek/EEM dwarf UBVIJHK colors Teff.txt}}. 
We then restrict the selection to stars located in the western Galactic hemisphere ($l \geq 180^{\circ}$, hereafter WGH), which is accessible to the SRG/eROSITA\_DE collaboration.

\begin{table}
\centering
\caption{Sample selection in this work based on \gaia\ 10 pc M- and F-, G-, K-type star (Stelzer, priv. comm. )}
\begin{spacing}{1.1}
\resizebox{\columnwidth}{!}{
\begin{tabular}{l|cc}
\hline
\toprule
Selection / Flag & M stars (N) & FGK stars (N)\\
\hline
\gaia\ 10 pc& 225 & 62\\
Western Hemisphere & 109 & 32  \\
\hline
Flag A: crowded field & 4 & 2 \\
\hline
Flag B: optical loading & 2 (excluded) & 22 (retained)\\
\hline
Flag C: close pair & 10 & 3 \\
\hline
Remaining   & 103   & 30  \\
(used in this work) & (Flag A|B = 0) & (Flag A = 0)\\

\hline
\end{tabular}}
\label{tab:sample}
\end{spacing}
\end{table}

\subsection{Sample overview}
Table~\ref{tab:sample} summarizes the number of stars for the sample selection. In total, the 10 pc-\gaia\ catalogue contains $225$ M0-M6 type stars and $57$ FGK type stars, of which,
109 M stars and 32 FGK stars are located in the WGH.

We remove 4 M stars (ID\,35, 45, 51, 208, index from Table~\ref{tab:M_alltable}) and 2 FGK stars (ID\,0 and ID\,43, index from Table~\ref{tab:FGK_alltable}) due to close proximity to other SpT X-ray sources (`crowded field' sources in Table~\ref{tab:sample}).
In addition, 2 M stars (ID 137, 157) were excluded from the final sample due to optical loading. 
We mark pairs of same SpT stars separated by less than $5^{\prime\prime}$ as `close pair's and extract a single spectrum from a common region encompassing both sources to avoid duplication (See Appendix~\ref{sec:exctraction} for details on the spectral extraction).
Eventually, 103 M stars remain after excluding the two stars affected by optical loading (see also Appendix~\ref{sec:opt_appdix}), constituting the final M dwarf sample used in this work. 
Instead, we do not exclude FGK stars affected by optical loading (22 of the 32), as excluding them would severely reduce the completeness of the sample.
The final FGK sample to be stacked still comprises 30 stars.
In Sect.~\ref{sec:optload} and Appendix.~\ref{sec:opt_appdix}, we will compare these 22 optically bright FGK stars' X-ray fluxes with the 9 optically faint FGK stars' and discuss how the impact of optical loading depends on optical brightness.

In Figure~\ref{fig:overview}, we present the eRASS1-detected \citep{Merloni2024AA} M-type and F-, G- and K-type stars that are used in our sample, showing X-ray luminosity versus \gaia\ colour (left panel) and total observed counts versus distance for all stars, regardless of whether they are detected in eRASS1 (right panel). 
Five FGK stars, $\alpha$ Cen A, $\alpha$ Cen B, Procyon A, {HD 156384 A and B}, were not shown in Figure~\ref{fig:overview}, as they lack \gaia\ measurements due to brightness saturation \cite{Reyle2021AA}.
The sample used in this paper has a wider SpT range for M stars (M0...M6) compared to \citet{Caramazza2023AA}, who have measured the X-ray luminosity of all the detected M0...M4 stars within 10 pc in both eastern and western hemispheres. 
According to the luminosity distribution, M stars (M0...M4) in the eastern Galactic hemisphere (EGH) appear to host more high luminosity sources than the western Galactic hemisphere (WGH).
Several of the most luminous M stars, such as AU Mic, AT Mic A, and GJ 867 A and B, are in fact located in the EGH.
On average, M stars in the EGH are about 4.7 times more luminous than those in the WGH.
For FGK stars (Stelzer B., in prep.), the average luminosity in the WGH is instead slightly higher than in the EGH, by a factor of about 1.3.
Therefore if one estimates the average luminosity from EGH, the result can be larger than the one we obtain in this work, as our data is limited to the WGH. 
In Appendix~\ref{sec:allstar}, we provide the identifiers, coordinates, distances, spectral types, the eRASS1 detection flag \citep{Merloni2024AA} and selection flag of the individual stars used in this work.

\subsection{Optical loading}\label{sec:optload}

In X-ray astronomy, optical loading refers to the interference caused by non-X-ray radiation, such as optical or ultraviolet light, on X-ray detectors \citep{Lumb2000}. While these detectors are primarily designed to detect X-rays, they can also exhibit sensitivity to other types of radiation, which can introduce noise and false signals, particularly in the soft X-ray regime \citep{Ramstedt2012AA, Ishikawa2019OpticalLS}.
Optical loading can artificially enhance the soft X-ray signal and potentially shift the measured energy of events across the entire spectrum.

To mitigate this effect, we adopt the \texttt{FLAG\_OPT} indicator from the eRASS1 catalogue \citep{Merloni2024AA}, which flags sources potentially affected by optical loading.
A source is assigned \texttt{FLAG\_OPT} = 1 if it is located within 15$^{\prime\prime}$ of a bright optical star (including the source itself) that meets any of the following brightness thresholds: B, V, or G $< 4.5$ mag, or J $< 3$ mag.
However, the optical loading threshold also depends on the survey depth: while G$<$4.5 mag was used for eRASS1, a threshold of G$<$5.0 mag is more appropriate for eRASS:4 (J. Robrade, priv. comm.).

\begin{figure}[]
    \centering
    \includegraphics[trim={0.cm 0.cm 0.0cm 0.3cm}, clip, width=\linewidth]{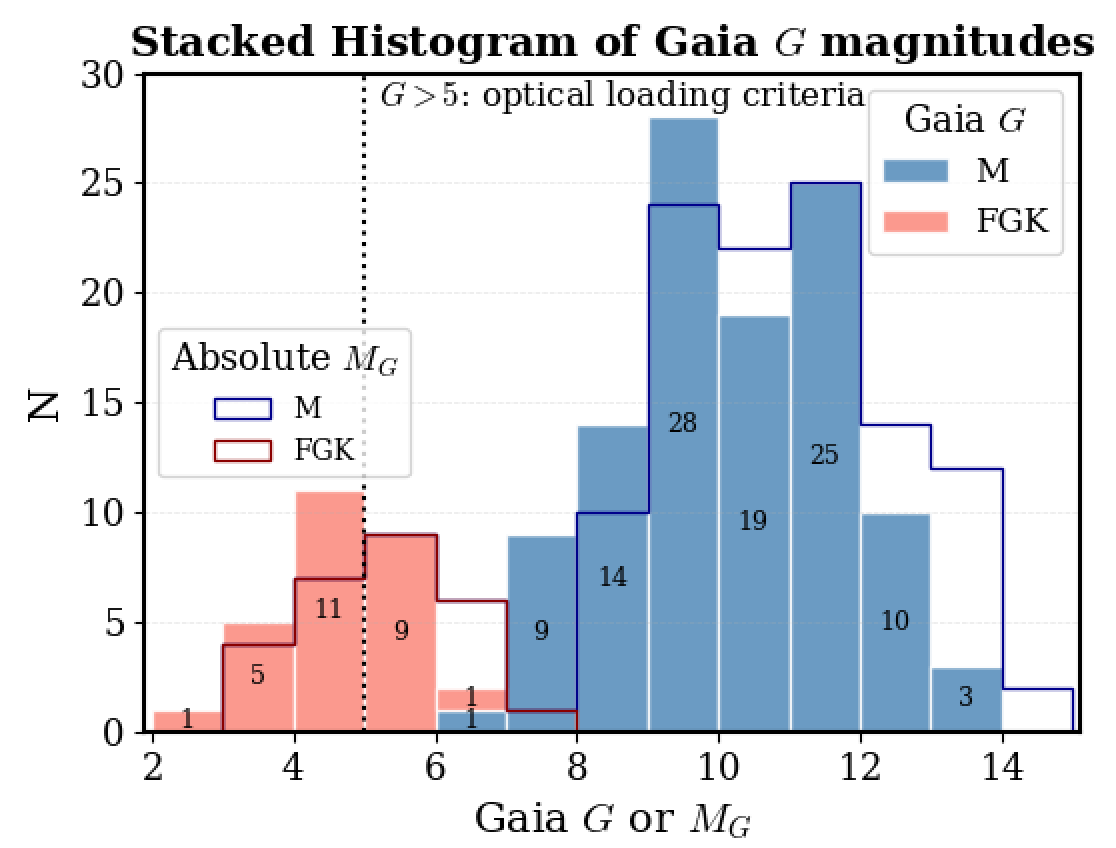}
    \caption{The histogram of the \gaia\ G apparent (G, filled bars) and absolute ($M_{G}$, outlined bars) magnitudes of the M-type (blue) and FGK-type (red) stars used in this study. The G mag and parallax is taken from \citet{Reyle2021AA}. Five stars with \gaia\ saturated magnitudes is not shown here. The vertical dashed line marks the optical loading threshold adpoted in this work is (G >5; J. Robrade, priv. comm.).}
    \label{fig:G_hist}
\end{figure}

In Figure \ref{fig:G_hist}, we can see no M stars in our sample suffers from optical loading (G < 5), in contrast to more than half of FGK stars. This result is expected, as the FGK stars are stronger optical emitters than M dwarf stars.
However, as noted earlier, two M dwarf stars were excluded due to optical loading: GJ 166 C based on its \texttt{FLAG\_OPT} indicator; GJ 442 B which is part of a visual binary with a separation of 22$^{\prime\prime}$ with the bright G-type primary star (GJ 442 A) affected by optical loading.

A detailed explanation of this effect, based on current knowledge of SRG/eROSITA and supported by spectral analysis, is provided in Appendix~\ref{sec:opt_appdix}.
Figure~\ref{fig:optload_m} compares M dwarfs and FGK dwarfs with different G-band magnitudes, showing that M stars spectra are consistent and confirming that optical loading has a negligible impact on the M dwarf sample. 
In contrast, for FGK stars, a similar test reveals differences on both luminosity and spectral shape, which could originate from optical loading effects (and the intrinsic luminosity scatter for this limited sample).
In Figures~\ref{fig01}–\ref{fig08}, we took three selected FGK stars ($G<5$) as example examining the possible impact of optical loading by comparing photon-event pattern selections (single, s) with the combined double+triple+quadruple selection (dtq).
To obtain a representative average luminosity, we retain all FGK stars but truncate the soft-energy range (< 0.35 keV) in this study. 

\section{Data processing}\label{sec:data}
The \erosita\ survey detects most of the late-type stars in 10pc-\gaia\ sample (about 75\% in eRASS1; \citealt{Merloni2024AA}, and over 90\% in eRASS:4, in prep.) and scanned all its stars four times over two years, from 2019 December 11 to 2021 December 19.
The eROSITA/eRASS:4 data were processed using the standard \erosita\ Science Analysis Software System (eSASS), version \texttt{eSASSusers\_240410}, developed by the German eROSITA consortium \citep{2022A&A...661A...1B}. 
For this study, we used event files from 5 telescope modules (\texttt{TMs}) with on-chips filters: \texttt{TM1,2,3,4,6}, the combination of which is referred to as \texttt{TM8}. 
Two modules (\texttt{TM5} and \texttt{TM7}) lack on-chip optical filters and suffer from light leaks \citep{Predehl2021AA}, so they were excluded from this study. 
Using the eSASS task \texttt{evtool}, event files from all four eRASS surveys were combined to create a single event set.
All the spectra are collected with all valid patterns (\texttt{PAT15}) when not explicitly mentioned.

\begin{figure}[]
    \centering
    \includegraphics[trim={0.cm 0.3cm 0.8cm 0.4cm}, clip, width=\linewidth]{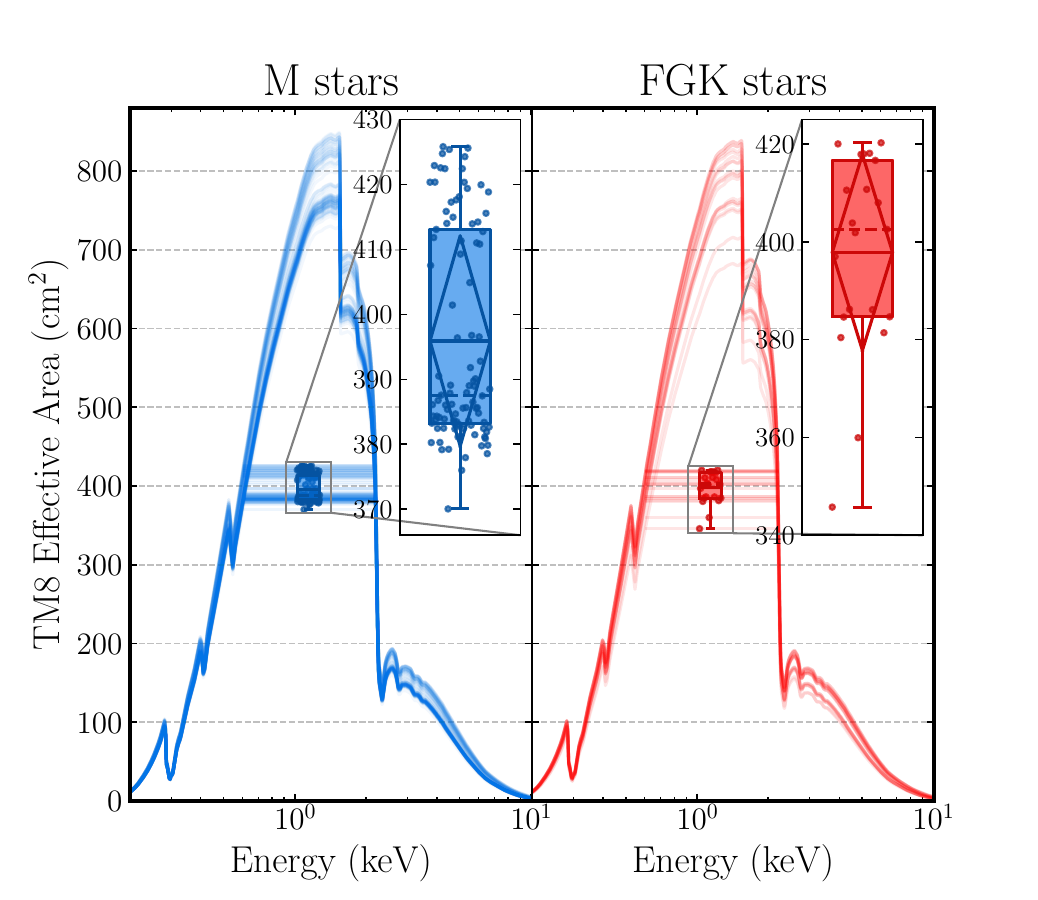}
    \caption{Comparison of the effective area of \texttt{TM8}. The box-plot represents the statistical distribution of the mean effective area over the 0.2--2.0 keV range. The solid horizontal line within each box indicates the average value, while the dashed horizontal line (in the zoomed-in plot) represents the median. The sample includes all 103 M stars and 30 FGK stars used in this study.}
    \label{fig:TM8_area_comparison}
\end{figure}

\section{Method: spectral stacking}\label{sec:method}

To stack the spectra of stars in the sample, the extraction region for each source was first adjusted to account for proper motion, aligning all positions to a standard reference epoch, December 15, 2020 \citep{eROsciencebook2012}.
A single spectrum was generated for each star by merging data from all four eRASS observations (see Appendix \ref{sec:exctraction} for further details on the extraction procedure).
In cases where two stars fell too close (within 25") in the same extraction region, only one combined spectrum was extracted, and both the source and background fluxes were scaled by a factor of two.
Each resulting spectrum was then normalized to a reference distance of 10 pc using a scaling factor $(d_i/10~{\rm pc})^2$ where $d_i$ is the distance to the star. The background subtraction was performed locally for each source.

Below, we present the full equation used to compute the average rates along with the associated uncertainty:
\begin{align}
C_{\text{net,total}} &= \sum_{i=1}^{N} \left[ \left(C_{\text{src},i} - C_{\text{bkg},i} \cdot S_i \right) \left( \frac{d_i}{10\,\text{pc}} \right)^2 \right]\label{eq:AC_C} \\
<R>&=\frac{C_{\text{net,total}}}{\sum_{i=1}^{N} T_i}\label{eq:AC_R} \\
<E_{R}>&=\frac{\sqrt{C_{\text{net,total}} }}{\sum_{i=1}^{N} T_i} = \sqrt{\frac{<R>}{\sum_{i=1}^{N} T_i}}\label{eq:AC_E}
\end{align}
$C_{\text{src},i}$ and $C_{\text{bkg},i}$ denotes the counts of source region and background from the $i$-th star, $<R>$ is averaged count rate of the $i$-th star, $d_i$ is its distance in pc, $T_i$ is the exposure time processed by \texttt{srctool}, $N$ is the total number of stars in the sample and $S_i$ is the scaling factor between sources region and background region. 
The uncertainties in the stacked spectrum $<E_R>$ reflect the combined Gaussian errors propagated from the individual source spectra.
Depending on the definition of stacking, there are multiple ways to combine spectra. The approach described in Eq.~\ref{eq:AC_R}, which we refer to as `Averaging Counts' (AC), is the method mainly used in this work. 
Since individual rate varies, the averaged rate can be more strongly influenced by stars with longer exposures.
While approximately 90\% of the stars have similar exposure times (processed by \texttt{srctool}) within one standard deviation, a few sources located near the ecliptic pole have significantly longer exposures and therefore might receive much higher weights. To ensure that these high-weight sources do not bias the average spectrum, we also performed additional stacking without them. A comparison of the results with and without these stars is presented in Appendix~\ref{sec:app_stacking}.

Alternatively, we could apply the `Averaging Rates' (AR) method for spectral stacking (Eq.\ref{eq:AR_R}-\ref{eq:AR_E}). Instead of summing total counts and dividing by total exposure time, this approach computes the average of the count rates across individual sources for each energy channel. Specifically, for each star, the count rate is distance-corrected and then averaged over all sources. 
The method is mathematically expressed as:
\begin{align}
    <R> &= \frac{\sum_{i=1}^{N}  (C_{\text{src},i} - C_{\text{bkg},i} \cdot S_i)/T_i \cdot \left(\frac{d_i}{10~pc}\right)^2}{N}\label{eq:AR_R} \\
    <C> &= <R> \cdot \sum_{i=1}^{N} T_i\label{eq:AR_C} \\
    <E_R>&= \frac{\sqrt{\sum_{i=1}^{N} (C_{\text{src},i} + C_{\text{bkg},i} \cdot S_i^2)\cdot (1/T_i \cdot \left(\frac{d_i}{10~pc}\right)^2)^2}}{N}\label{eq:AR_E}
\end{align}
where the denotation follows the same as Eq.\ref{eq:AC_C}-\ref{eq:AC_E}.
This approach normalizes counts by exposure before averaging, which ensures each source contributes according to its count rate. 
However, because standard error propagation assumes statistical independence, it can lead to overestimated uncertainties when stacking spectra with correlated measurements or shared systematics.
To mitigate this, we adopt a Gaussian noise assumption, which provides a lower bound on the uncertainties, acknowledging that this likely underestimates the actual statistical error.
In Appendix~\ref{sec:app_stacking}, we compare the resulting fits and luminosities from this method with those obtained using the AC approach. The differences between them are treated as a source of systematic uncertainty in the final luminosity estimates.

In Appendix~\ref{sec:app_stacking}, we compare the best-fit parameters and luminosities obtained from the AC and AR methods, as well as when excluding highly exposed sources, confirming that the stacking procedure does not significantly affect the overall spectral shape.

\begin{figure*}
    \centering
    \includegraphics[trim={0.6cm 0.1cm 0.1cm 0.1cm}, clip, width=0.95\linewidth]{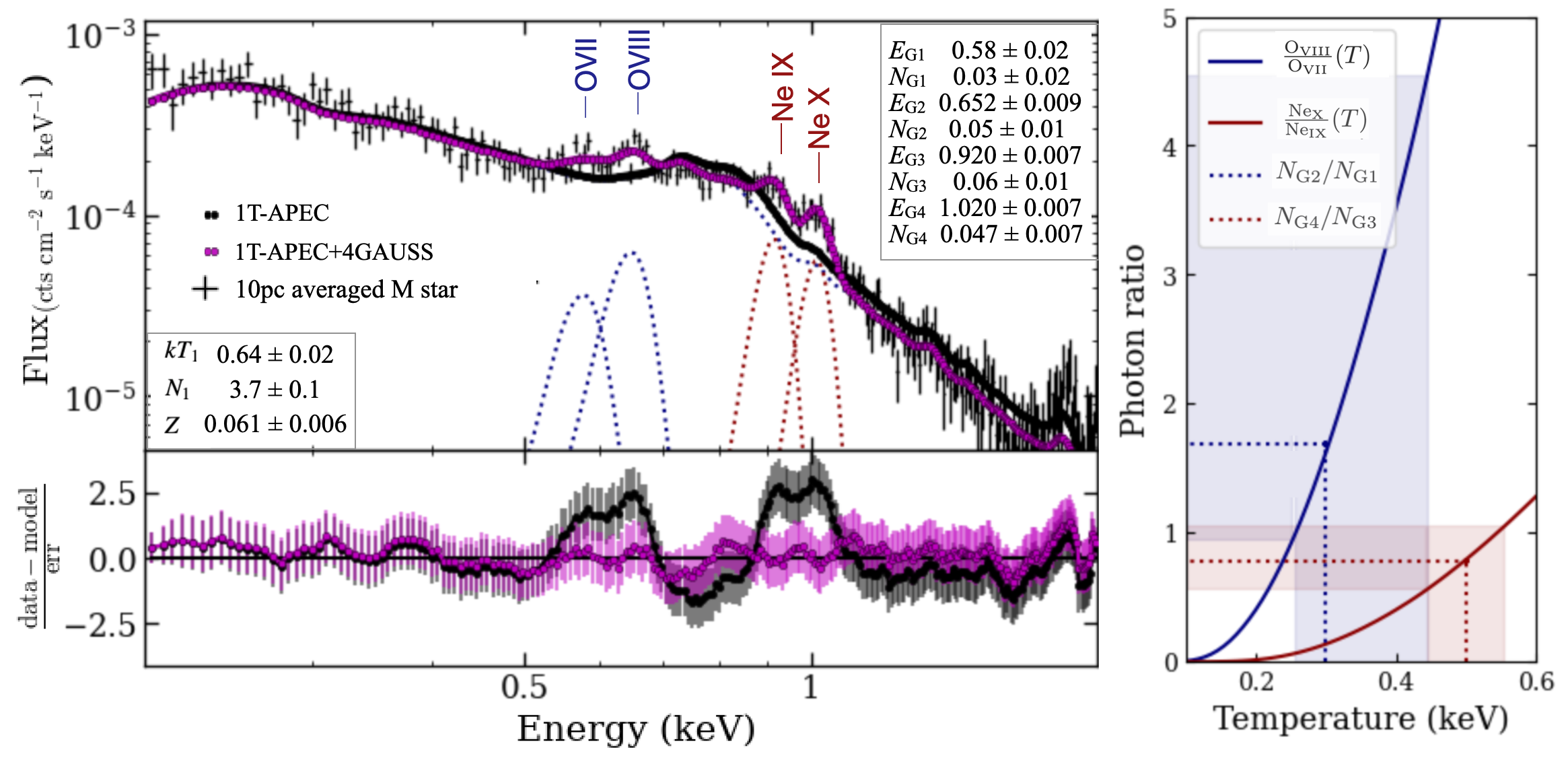}
    \caption{Left panel: Averaged 10 pc M dwarf spectrum with different models. All the models are labeled in the legend, where names follow the \texttt{XSPEC} convention, and the letter G stands for the Gaussian components to fit single emission lines (dotted lines). 
    Right panel: temperature vs. line ratio assuming a single thermal component. The dark red and navy lines show the expected $\ion{O}{VIII}/\ion{O}{VII}$ and $\ion{Ne}{X}/\ion{Ne}{IX}$ ratios from \texttt{AtomDB}. Vertical dashed and dotted lines mark ratios derived from Gaussian normalizations in Table~\ref{tab:Mfit}.}
    \label{fig:M_spec_fit_apec}
\end{figure*}

\begin{figure*}
    \centering
    \includegraphics[trim={0.2cm 0.2cm 0.1cm 0.0cm}, clip, width=0.95\linewidth]{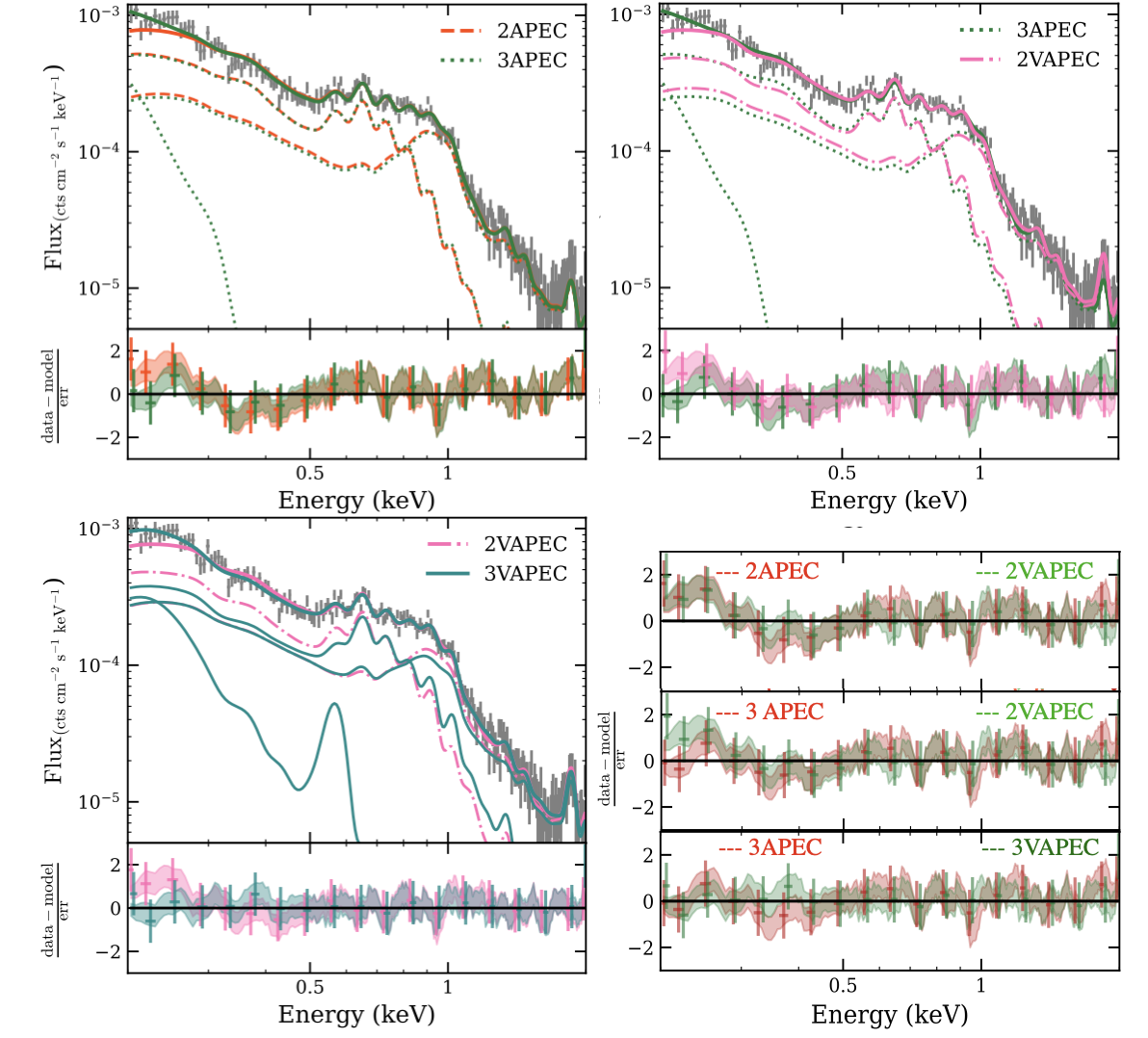}
    \caption{
Model fitting of the averaged X-ray spectra (0.2--2.0~keV) of 103 M dwarf stars using multi-temperature thermal plasma models. The top-left and top-right panels show comparisons between 2-temperature and 3-temperature APEC models (labeled as 2APEC and 3T-APEC), and as variable-abundance counterparts (2T-VAPEC and 3T-VAPEC). The corresponding residuals are also shown. The bottom-left panel displays the spectral fits for 2T-VAPEC and 3T-VAPEC models, highlighting differences in spectral shape and emission features. The bottom-right panel presents a comparative view of residuals amongst all four models to assess the fit quality. Residuals are shown in standard deviations (in sigmas).}\label{fig:M_spec_fit_4model}
\end{figure*}

Given that the effective area of the detector varies with vignetting and masking in the source region, an averaged effective area was computed in each PI channel. While this method is not theoretically optimal, it provides a practical and reproducible approach. 
Figure~\ref{fig:TM8_area_comparison} presents a statistical comparison of the effective area of \texttt{TM8} for the sample of 103 M stars and 30 FGK stars. 
The box plot displays the distribution of the mean effective area. 
In each box, the solid horizontal line represents the mean value of the sample, while the dashed line indicates the median value. 
Variations in the effective area amongst sources were mainly caused by masking nearby point sources, with the maximum discrepancies reaching approximately 10\% for both samples, as illustrated in the box-plot.

\section{Analyse: spectral fitting} \label{sec:abundance}
The spectral fitting is performed using the \texttt{pyXspec} software \citep{Arnaud1996ASPC}. The fitting process assumes elemental abundances from \citet{Asplund2009ARAA} and photoelectric absorption cross-sections from \citet{Verner1996ApJ}.
$\chi^2$ minimization is used for the fitting. 

Our analysis assumes that the emitting plasma is in collisional ionization equilibrium (CIE) \citep{Raymond1977ApJS, Gudel2004AA}.
The analysis begins with a single-temperature collisionally ionized plasma model, APEC (Astrophysical Plasma Emission Code; \citet{Smith2001ApJ}) model, which describes a collisionally ionized plasma in stellar coronae of thermal equilibrium. 
It then progresses towards more complex models, including multi-temperature (2T-APEC, 3T-APEC) \citep{Smith2001ApJ} and variable abundance models (1T-VAPEC, 2T-VAPEC, and 3T-VAPEC).
The elemental abundances are tied across all thermal components.

We denote the overall metallicity by $Z$ and the abundance of a specific element by $A_X$.
The metal abundance ($A_X$) is defined as $A_{\rm element}=\frac{(n_{Z}/n_{H})_{\rm plasma}}{(n_{Z}/n_{H})_{\odot}}$.
With VAPEC model, we allow O, Ne, Si, and Fe to vary independently, all other elements (He, C, N, Mg, Al, S, Ar, Ca, Ni) are tied together and allowed to vary as a single free parameter.
Alpha-elements such as oxygen, neon, and silicon are primarily considered \citep{Bensby2014AA, Jofre2015AA, Montes2018MNRAS}.
We verified that we do not obtain improvement in the fit by untying any of the abundances of the other elements.
More parameter initializations are listed in Appendix~\ref{sec:morevapec}.

To assess the relative quality of spectral fits, we employed the Fisher test (F-test), 
the Bayesian Information Criterion (BIC; \citealt{Schwarz1978AnSta}), and the Akaike Information Criterion (AIC; \citealt{Akaike1974ITAC}). The F-test evaluates whether the addition of extra parameters significantly improves the fit, assuming nested models and Gaussian errors. The AIC and BIC are both information-theoretic criteria that balance model goodness-of-fit against complexity, with BIC applying penalty for the complex models. In our case, we adopt AIC=$2k+\chi^2_\mathrm{stat}$ and BIC=$k~\mathrm{ln}(N)+\chi^2_\mathrm{stat}$, where k is the number of free parameters and N is the number of data channels.

\section{Results: M dwarfs sample}\label{sec:model_fitting}

In this section, we fit the average X-ray spectrum of M dwarfs. 
Unless stated otherwise, the M star spectra are fit within the energy range of 0.2--2.0\,keV, where the S/N of the stacked spectra, which is scaled to 10 pc, exceeds 1.

The results of the fitting procedure for the sequence of models are summarized in Table~\ref{tab:Mfit}, while the corresponding multi-temperature spectral fits are presented in Figure~\ref{fig:M_spec_fit_4model}.
The uncertainties correspond to 90\% confidence intervals for a single parameter assuming chi-square statistics.
The bootstrap uncertainties (90\% CI) are reported as the second term in the brackets of the corresponding parameter.
The F-test between models are provided in Table~\ref{tab:model_ftest_comparison}.

\begin{table*}[]
\caption{Model parameters and values for M dwarf averaged spectrum.}
\begin{spacing}{1.4}
\centering
\resizebox{\textwidth}{!}{
\begin{tabular}{ l | c c|c c |c c| c c}
\hline
\toprule
Param.	&	APEC	&	VAPEC	&	2T-APEC	&	2T-VAPEC	&  3T-APEC	&	3T-VAPEC&		&	APEC +GAUSS\\
 \hline
$kT_1$	&	$0.66 \pm 0.02$	&	$0.46\pm0.02$&	$0.27~(\pm 0.01, \pm0.01)$	&	$0.26~(\pm0.02, {}^{+0.04}_{-0.01} )$&	$0.10~(\pm0.10,{}^{+0.01}_{-0.09})$&	$0.09\pm0.09$	&	$kT_1$	&	$0.60 \pm 0.02$\\
$N_1$	&	$3.7 \pm 0.1$	&	$4.9 \pm0.3$&	$1.7~(\pm 0.1,^{+0.1}_{-0.3})$	&	$2.7~(\pm 0.6, {}^{+0.1}_{-0.7})$	&	$0.3~({}^{+37.3}_{-0.2} ,{}^{+180.0}_{-0.1})$&	$0.7^{+0.5}_{-0.6}$	&	$N_1$	&	$3.7\pm0.2$\\
$Z~{\rm or}~A_{\rm Rest}$	&	$0.061 \pm 0.006$	&	$0.10 \pm 0.04$&	$0.28~(\pm 0.04, {}^{+0.08}_{-0.01})$	&	$0.15~({}^{+0.11}_{-0.08},{}^{+0.05}_{-0.08})$	&	$0.31~(\pm0.06,  {}^{+0.70}_{-0.01})$&	$0.143^{+0.08}_{-0.05}$	&	$Z$	&	$0.048 \pm 0.006$\\
$kT_2$	&		&		&	$0.94~( \pm 0.03,{}^{+0.05}_{-0.01})$	&	$0.96~(\pm0.05, \pm0.04)$	&	$0.27~(\pm 0.02,\pm0.07)$&	$0.26^{+0.03}_{-0.06}$	&	$E_{G1}$	&	$0.58\pm0.02$\\
$N_2$	&		&		&	$1.1~(\pm 0.1,{}^{+0.1}_{-0.2})$	&	$1.6~(\pm0.3,{}^{+0.1}_{-0.3})$	&	$1.5~(\pm 0.2,{}^{+0.1}_{-0.2})$&	$1.5^{+0.5}_{-0.7}$	&	$N_{G1}$	&	$0.03\pm0.02$\\
$kT_3$	&		&		&		&		&	$0.95~(\pm 0.03,{}^{+0.05}_{-0.01})$&	$0.63^{+0.09}_{-0.04}$	&	$E_{G2}$	&	$0.652\pm0.009$\\
$N_3$	&		&		&		&		&	$1.0~(\pm 0.1,{}^{+0.1}_{-0.2})$&	$2.6^{+0.4}_{-0.8}$	&	$N_{G2}$	&	$0.05 \pm 0.01$\\
$A_\mathrm{O}$	&		&	$0.12^{+0.04}_{-0.03}$&		&	$0.16~(\pm 0.06, {}^{+0.08}_{-0.01})$	&		&	$0.18_{-0.05}^{+0.09}$	&	$E_{G3}$	&	$0.920\pm0.007$\\
$A_\mathrm{Ne}$	&		&	$0.47 \pm 0.03$&		&	$0.2~(\pm 0.1, {}^{+0.2}_{-0.1})$	&		&	$0.7 \pm 0.2$	&	$N_{G3}$	&	$0.06 \pm 0.01$\\
$A_\mathrm{Si}$	&		&	$0.3 \pm 0.1$&		    &	$0.1~(\pm 0.1,{}^{+0.2}_{-0.1})$	&		&	$0.18 \pm 0.10$	&	$E_{G4}$	&	$1.020\pm0.007$\\
$A_\mathrm{Fe}$&		&	$0.041\pm0.008$&		&	$0.20~(\pm0.02, {}^{+0.02}_{-0.06})$	&		&		$0.07\pm0.02$&	$N_{G4}$	&	$0.047 \pm 0.007$\\
\addlinespace[2pt]
     \hline
$\chi_{\nu}^2\ (\frac{\chi^2}{\mathrm{d.o.f.}})$	&	$2.12
(\frac{531.65}{251})$	&	$1.03(\frac{255.33}{247})$&	$0.93(\frac{231.50}{249})$	&	$0.92(\frac{225.19}{245})$	&	$0.91(\frac{224.58}{247})$	&	$0.84(\frac{202.24}{243})$	&	$\chi_{\nu}^2\ (\frac{\chi^2}{\mathrm{d.o.f.}})$	&	$0.93(\frac{225.16}{243})$\\
AIC	&	537.65	&	269.33 &	241.50	&	243.19	&	238.58	&	224.24	&	AIC	&	247.16\\
BIC	&	548.26	&	294.09 &	259.20	&	275.03	&	263.34	&	264.15   &	BIC	&	286.07\\
\bottomrule
\end{tabular}}
\end{spacing}
\tablefoot{Normalization $N_{\rm apec}$ is in the unit $10^{-4}\frac{{10}^{-14}}{4\pi(10 {\rm pc})^2}~ \mathrm{cm}^{-5}$. The set of abundance is bonded for all components of a model. The error reported corresponds to 90 \% confidence, assuming chi-square statistics. In model 2T-APEC, 2T-VAPEC, and 3T-APEC, the second error is the 90\% confidence interval from the bootstrap resampling distribution (Section~\ref{sec:bootstrap}). $A_{\rm element}=\frac{(n_{Z}/n_{H})_{\rm plasma}}{(n_{Z}/n_{H})_{\odot}}$ . The 2T-APEC model is selected to model the flux.}\label{tab:Mfit}
\end{table*}

\subsection{Single temperature plasma models}\label{sec:1component}
The one-temperature APEC model yields a poor fit  ($\chi_\nu^2 = 2.12$) with $kT = 0.66 \pm 0.02$ keV and an unusually low abundance of $Z_{\rm all} = 0.061 \pm 0.006 Z_{\odot}$, 
 lower than the typical values found in stellar coronae, $Z\approx 0.3-0.4\, Z_\odot$ (e.g. \citep{Favata2004AA}, and \citep{Robrade2005AA}).
 
This is likely a spectral modeling bias related to stacking, similar to the `artificially Fe bias' in galaxies group and cluster studies when a multi-temperature plasma (with different thermal components) is modelled using a single-temperature APEC model \citep{Buote2000ApJ, Sanders2011MNRAS}.

Residuals at 0.5–0.7 keV and 0.9–1.1 keV align with known strong emission lines, such as O VII, O VIII, Ne IX and Ne X. This is demonstrated by the better fit obtained after adding four Gaussian profiles that fit energies approximately equal to the ones of the brightest emission lines. 
In addition, by interpreting the normalization of the added Gaussian profiles as intensities of a plasma emission, the ratio between the normalizations of the profiles corresponding to O VIII/O VII (purple dashed lines) and Ne X/Ne IX (purple dotted lines) does not match the expectation of a single temperature plasma, as shown by the right panel of Figure~\ref{fig:M_spec_fit_apec}. The theoretical line ratios are independent of the relative abundances of the elements. 
The fact that we derive different ratios for O and Ne strongly hints at the need for an additional temperature component.

\subsection{Multi-temperature plasma models}

A dual-temperature APEC model (2T-APEC: $\chi_\nu^2= 0.93$) provides a significant improvement over the single-temperature APEC model and performs better than the single VAPEC model, as indicated by the F-test (see Table~\ref{tab:model_ftest_comparison}) as well as by the AIC and BIC criteria.
This suggests that the improvement in the fit is primarily driven by the inclusion of a second thermal component rather than by adjustments to the elemental abundances.
The two plasma temperatures are $kT_1 = 0.27\pm0.01$ keV and $kT_2 = 0.94\pm0.03$ keV, with a tied abundance of $Z_{\rm all} = 0.28\pm0.04$. 
Though it is a relatively good fit, the residuals shown in the upper left panel of Figure~\ref{fig:M_spec_fit_4model} show the presence of a significant excess below $0.3$\,keV and a bump near 1.9 keV.

Adding a third temperature component (3T-APEC: $\chi_\nu^2= 0.91$) further improves the fit (see F-test in Table~\ref{tab:model_ftest_comparison}; AIC and BIC in Table~\ref{tab:Mfit}).
The additional cooler component has poorly constrained temperature and normalization: $kT_1 = 0.1\pm0.1$ keV and $N_1=0.3^{+37.3}_{-0.2}$.
From the upper left panel in Figure~\ref{fig:M_spec_fit_4model}, we can see this component contributes only a tiny range of energies in the spectrum (<0.3 keV), effectively coming up (only) for the low-energy excess seen in the 2T-APEC model.
In fact, the fitted temperatures of the second and third components are consistent with those obtained in the 2T-APEC model. 
Adding this component yields only a modest improvement, and the F-test does not support it as statistically significant (p-value=0.024).

\begin{table}[h]
    \caption{F-test statistics of models for average M star.}
    \centering
    \resizebox{\columnwidth}{!}{
    \begin{tabular}{cc|cccc}
        \hline
        \toprule
Simpler	&	More~complex	&	\multirow{2}{*}{$\Delta$ d.o.f.}	&	\multirow{2}{*}{$\Delta \chi^2$}	&	\multirow{2}{*}{F-value}	&	\multirow{2}{*}{p-value}	\\
Model & Model & & & & \\
\hline							
\addlinespace[2pt]
1T-APEC	&	2T-APEC	&	2	&	300.15	&	161.4	&	$1\times10^{-45}$	\\
2T-APEC	&	3T-APEC	&	2	&	6.92	&	3.8	&	$ 0.024$	\\
\hline												
1T-APEC	&	1T-VAPEC	&	4	&	276.32	&	66.8	&	$3\times10^{-38}$	\rule{0pt}{10pt} \\
2T-APEC	&	1T-VAPEC	&	2	&	-23.83	&	<0	&	-	\\
2T-APEC	&	2T-VAPEC	&	4	&	23.69	&	1.7	&	0.14	\\
3T-APEC	&	2T-VAPEC	&	2	&	-0.61   &	<0	&	-	\\
3T-APEC	&	3T-VAPEC	&	4	&	22.34	&	6.7	&	$4\times10^{-5}$	\\
\hline													
1T-VAPEC	&	2T-VAPEC	&	2	&	30.14	&	16.4	&	$2\times10^{-7}$	\rule{0pt}{10pt}\\
2T-VAPEC	&	3T-VAPEC	&	2	&	22.95	&	13.8	&	$2\times10^{-6}$	\\
     \bottomrule
    \end{tabular}}
    \tablefoot{The improvement in best-fit between two models regarding change in degrees of freedom ($\Delta$ d.o.f.), change in $\chi^2$, the corresponding F-value, and the resulting p-value. The d.o.f. is $N_\mathrm{data}-N_\mathrm{free~param.}$.}
    \label{tab:model_ftest_comparison}
\end{table}

We further applied 2T- and 3T-VAPEC models, with the elemental abundances linked between the two or three components. 
As mentioned in Sect.~\ref{sec:abundance}, the abundances of element O, Ne, Si, and Fe were allowed to vary freely, while all other elements are tied together to vary.
According to the statistical estimators, the 2T-VAPEC model ($\chi_\nu^2= 0.92$) provides a comparably good fit to the 2T-APEC model, but requires four additional free parameters, making it the less favoured choice.

The 3T-VAPEC model ($\chi_\nu^2= 0.84$) appears mildly over-fitted, requiring an additional third component at a very soft temperature of \(0.09\pm0.09\,\mathrm{keV}\), which is not constrained at all. 
Varying the elemental abundances alone does not resolve the soft excess and only shuffles the relative strength of the elements/components of the fit (See Appendix.\ref{sec:morevapec}). 

This poorly constrained very soft component is atypical for M stars and may instead reflect calibration uncertainties at the lowest energies. After updating to the most recent response files, calibrated using soft neutron-star spectra, its significance is already reduced than previous version.
We therefore restrict the subsequent analysis to two-temperature models and adopt the 2T-APEC model as our baseline. This model provides a simple description while retaining the key feature that its two components at 0.27 keV and 0.96 keV are also consistently required in the 2T-VAPEC and 3T-APEC models. We use the 2T-APEC model as the best fit for the following subgroup analysis of the M-star sample.

\subsection{Luminous M dwarfs versus the remaining sample}\label{sec:brightM}

\begin{table}[hb!]
\caption{Fitting parameters for different luminosity groups. }\label{tab:Mdwarf_ADLeo2}
    
    \begin{spacing}{1.2}
    \centering
    \resizebox{\columnwidth}{!}{
    \begin{tabular}{l | cccc}
        \hline
        \toprule
\multirow{2}{*}{Param.}	&	\small 99 lower &102 lower	& 4 higher	& AD Leo	\\
&	luminosity & luminosity	&  luminosity	& (at 10~pc)\\
\hline					

$kT_1$	&	${ 0.27}\pm0.02$ & $0.27\pm0.01$	& ${ 0.28}\pm0.02$	& $0.26\pm0.01$	\\
$N_1$	&	${ 1.3\pm0.1}$   & ${ 1.5\pm0.1}$	&	${ 20\pm3}$ &  ${ 25}\pm5$	\\
$kT_2$	&	${ 0.92}\pm0.04$ & $0.93\pm0.03$ & ${ 1.03^{+0.05}_{-0.06}}$ & ${ 0.94^{+0.08}_{-0.11}}$	\\
$N_2$	&	${ 0.8}\pm0.1$   & ${ 1.0}\pm0.1$	& ${ 19\pm4}$	& ${ 16^{+7}_{-5}}$	\\
$kT_3$	&	   &    & & ${ 1.5^{+0.7}_{-0.3}}$	\\
$N_3$	&	   &  	& & ${ 15^{+7}_{-5}}$	\\
$Z/Z_\odot$	&  ${ 0.24^{+0.05}_{-0.04}}$ &	${ 0.26^{+0.05}_{-0.04}}$	&	${ 0.29^{+0.09}_{-0.07}}$ & $0.5^{+0.2}_{-0.1}$	\rule[-5pt]{0pt}{2pt} \\
\hline					
$\chi^2_\nu$ & $0.6~(\frac{{ 132.73}}{210})$	& $0.8~(\frac{{ 178.69}}{224})$	& $0.4~(\frac{{ 83.24}}{202})$	&$1.3~(\frac{{ 255.12}}{201})$	\rule[-0pt]{0pt}{10pt}\\
$L_\textrm{0.2-2.5}$	& $1.81\pm{ 0.05}$	& ${ 2.26\pm0.05}$	& ${ 39\pm1}$ & ${ 73\pm2}$	\\
     \bottomrule
    \end{tabular}}
\tablefoot{Fitting is done for spectra in the 0.2--2.5 keV range. 3T-APEC model is used for AD Leo as it is significantly required. All spectra are normalized to 10 pc and binned to a minimum of 10 counts per channel. $N_{\rm apec}$ is in unit of $10^{-4}\frac{{10}^{-14}}{4\pi(10 {\rm pc})^2}~ \mathrm{cm}^{-5}$. Luminosities are given in $10^{27}\mathrm{erg/s}$ unit. 90\% chi-square statistics error is shown.}
    \end{spacing}
    \end{table}

\begin{figure*}
    \centering
    \begin{subfigure}{0.48\textwidth}
    \includegraphics[width=\linewidth]{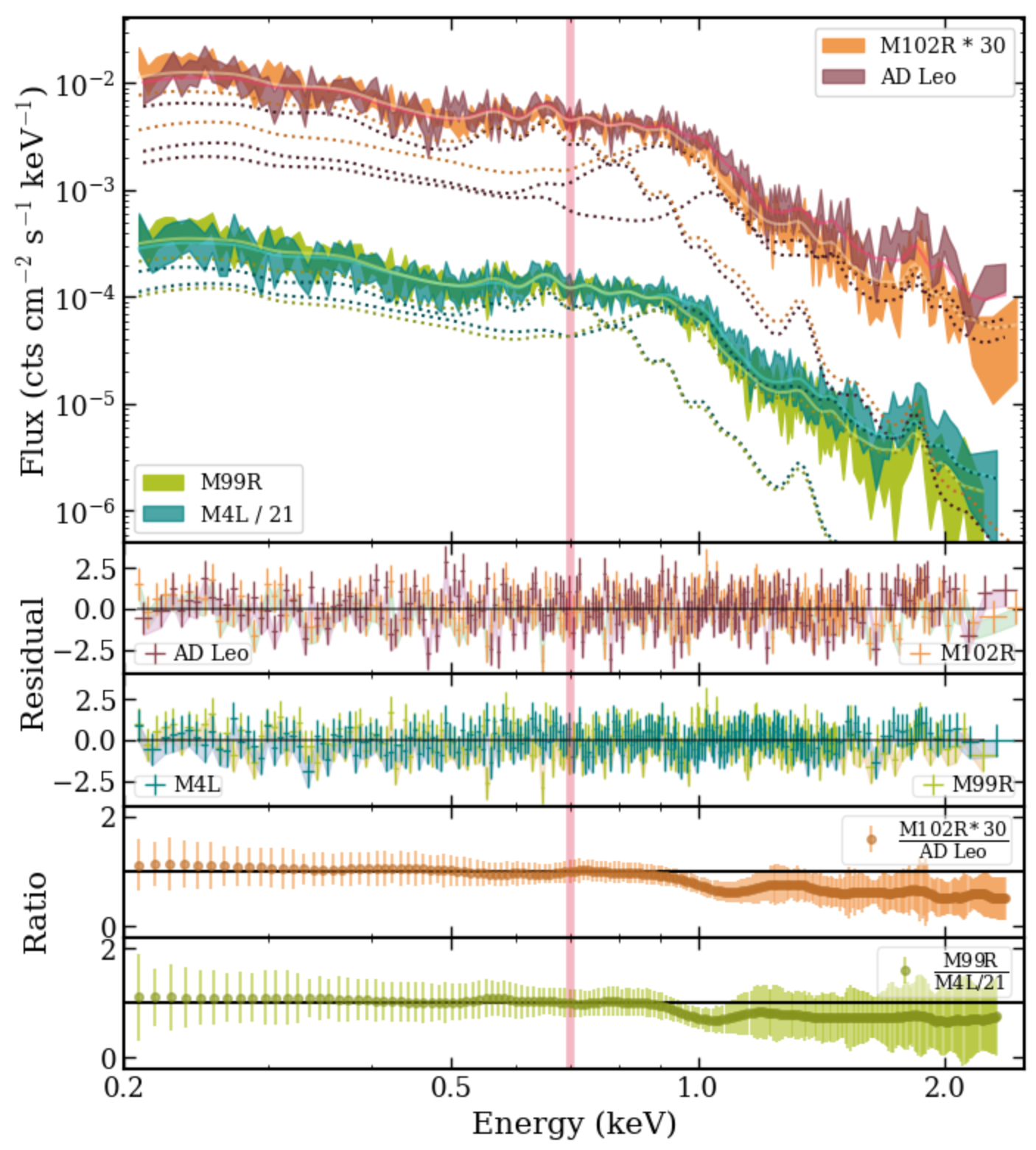}
    \caption{AD Leo (brown) and the average fainter 102 M stars (M102R: orange), bright 4 M stars (M4L: blue) and fainter 99 M stars (M99R: green). 0.2--2.5 keV range is considered for model fitting.}\label{fig:adleo}
    \end{subfigure}
    \hfill
        \begin{subfigure}{0.48\textwidth}
    \includegraphics[width=\linewidth]{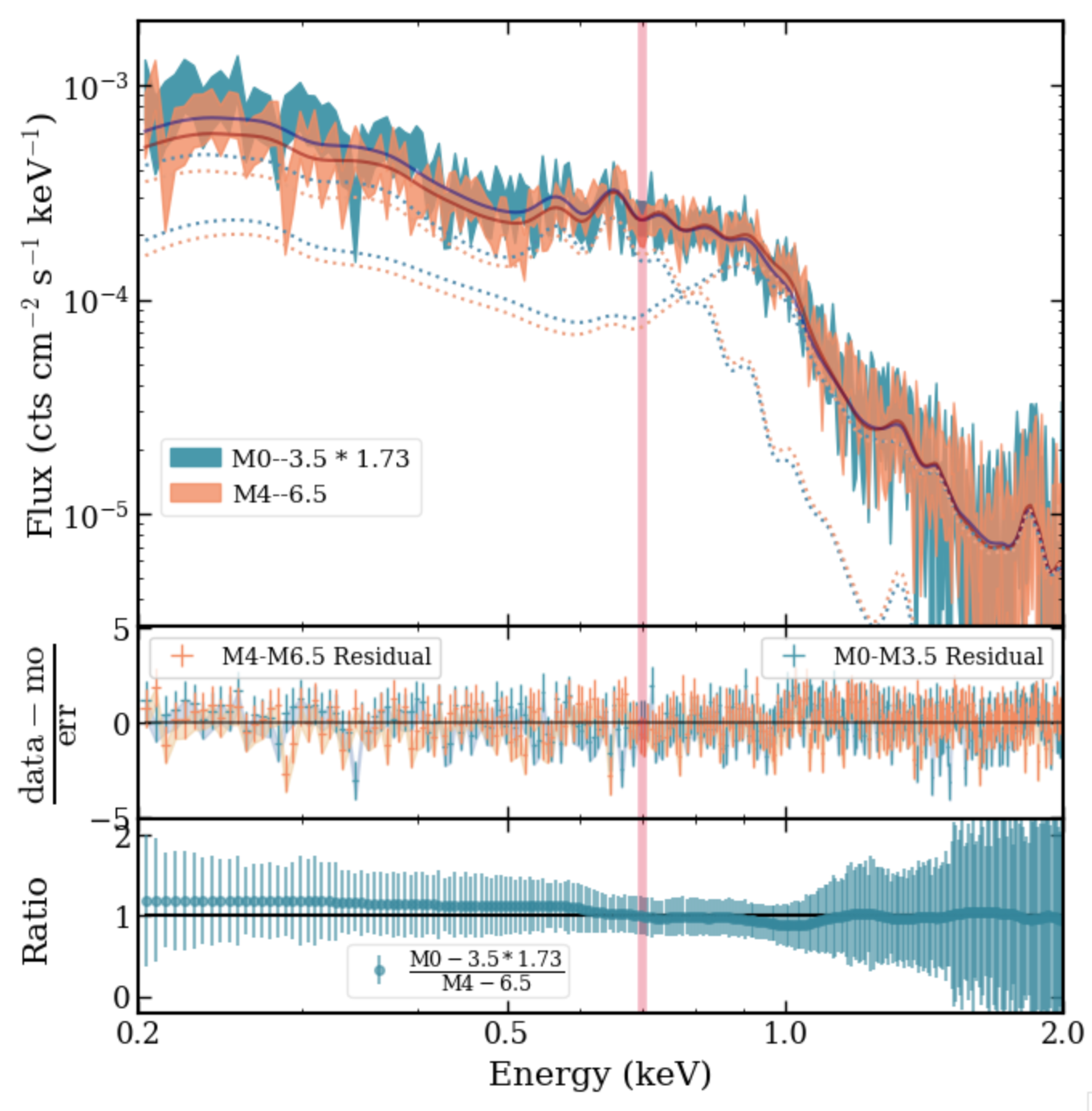}
    \caption{Average spectra of 57 M early-type stars (M0--M3.5: blue) and the 44 M mid-late-type stars (M4--M6.5: orange). 0.2--2.0 keV range is considered for model fitting.                           }\label{fig:M0346}
    \end{subfigure}
    \caption{Comparison between X-ray brightness subgroups and subspectra type groups. All spectra are normalized to a distance of 10 pc and aligned at 0.7 keV for comparison. Each spectrum is fitted with a 2T-APEC model. The residual ($\frac{\rm data-model}{\rm error}$) from fit and the ratio between the two spectra is shown.}
\end{figure*}

\begin{table}
    \caption{Fitting parameters for average spectra of M0–M3.5 and M4–M6.5 subgroups. }
    \centering
    {\begin{spacing}{1.25}
    \begin{tabular}{l | cc}
        \hline
        \toprule
Param.	& [M0-M3.5]	&	[M4-M6.5]	\\
\hline					
$kT_1$	&	${ 0.26\pm0.02}$	&	$0.27\pm0.01$\\
$N_1$	&	${ 1.5}\pm0.2$  	&	${ 1.9\pm0.3} $	\\
$kT_2$	&	${ 0.92}\pm0.05$	&	${ 0.96}\pm0.04$	\\
$N_2$	&	${ 0.9\pm0.2}$	    &	${ 1.3}\pm0.2$	\\
$Z/Z_\odot$	&	${ 0.24^{+0.06}_{-0.05}}$	&	{ $0.34^{+0.08}_{-0.06}$}	 \\
\hline					
$\chi^2_\nu (\frac{\chi^2}{\rm d.o.f.})$	&	${ 0.82~(\frac{205.21}{249})}$	&	${ 0.74~(\frac{183.69}{249})}$	\\
$L_\textrm{0.2-2.0}$	&	{ $2.03\pm0.06$}	&	{ $3.34\pm0.09$}	\\
     \bottomrule
    \end{tabular} 
    \end{spacing}}
    \tablefoot{Fitting is done for spectra in the 0.2--2.0 keV range. Same treatment and denotation as Table~\ref{tab:Mdwarf_ADLeo2} is applied.}
    \label{tab:M0346}
\end{table}

 The X-ray luminous stars potentially skews the spectral characteristics from the average spectra shape. 
In this section, we examine how AD Leonis and other very luminous stars affects the features of stacked average spectra.
As shown in the left panel of Figure \ref{fig:overview}, AD Leo (AD Leo or BD+20 2465) stands out as the most luminous in our sample, lying near the upper envelope of the luminosity distribution. 
Although its colour corresponds to an early-M type (M3.5V), its X-ray luminosity is nearly an order of magnitude higher than that of typical M0--M4 stars, consistent with its well-known status as a magnetically active flare star \citep{Stelzer2022AA}.
 Following AD Leo are YZ CMi and G 41-14 AB, which are themselves about 2 times more luminous than AP Col. We therefore define these four brightest M stars as group M4L group.
 Hence we define 2 groups of stars: AD Leo to the remaining 102 M dwarfs with lower luminosity (M102R), and the four most luminous M stars (M4L) with the remaining 99 less luminous M stars (M99R).

Figure~\ref{fig:adleo} shows the average X-ray spectra of the four groups, with the best-fit models from Table~\ref{tab:Mdwarf_ADLeo2} overlaid.
The 2T-APEC model is applied to the data over the 0.2--2.5 keV, a broader range than in the other analysis of this work, to examine the harder spectrum that is expected for active stars.
The average spectrum of M102R is scaled up by a factor of 30 to match the distance-normalized flux of AD Leo at 0.7 keV, while the M4L spectrum is scaled down by a factor of 20 to align with the flux of M99L at the same energy.
From the ratio panel in the bottom of Figure \ref{fig:adleo} (flux normalized at 0.7 keV), we observe that AD Leo closely resembles the M102R in the intermediate energy range (0.3–0.9 keV). 
The deviations rise lower energies (< 0.3 keV) and higher flux at higher energies (> 0.9 keV) as AD Leo exhibits overall harder spectrum.
A similar trend is seen for M4L when compared with M99R.

The best fitting parameters are listed in Table \ref{tab:Mdwarf_ADLeo2}. 
The average spectra of M99R, M102R and M4L are well described by two components with temperatures of $\sim$0.27 and 0.92--1.03 keV, yielding good fits with $\chi_\nu^2 <$ 0.8 that indicate slight over-fitting.
The fit for spectrum of AD Leo are fitted with components at 0.26, 0.94, and 1.5 keV. The fitting of AD Leo is marginally acceptable as $\chi_\nu^2$=1.3. The residuals scatter at 0.45--0.6 keV.

The M102R, M99R and M4L samples have remarkably consistent best-fitting parameters ($T$ and $Z$) with each other and with the whole M star sample in Table~\ref{tab:Mfit}.
All spectra share a consistent intermediate-temperature component ($\sim$0.27 keV) and a hotter components at ($\sim$0.9 keV). Those two contribute the majority of the emission. 
Only AD Leo exhibits an additional high-temperature component around $1.5$ keV and higher coronal abundances ($0.5^{+0.2}_{-0.1}$) compared to the rest groups.
The relatively poor fit of AD Leo may reflect its distinct magnetic activity that are not fully captured at the current energy resolution.
Despite its individual peculiarities, either AD~Leo's and M4L's overall spectral shapes remain broadly similar to that of other M dwarfs in the central energy range.
By comparing with the best-fit of all M stars in Table~\ref{tab:Mfit} we conclude that, either excluding AD~Leo or the most luminous 4 stars from the stacked sample, does not significantly alter any parameter except for the overall normalization.

\begin{figure*}
    \centering
    \includegraphics[trim={0.2cm 0.2cm 0.1cm 0.0cm}, clip, width=0.95\linewidth]{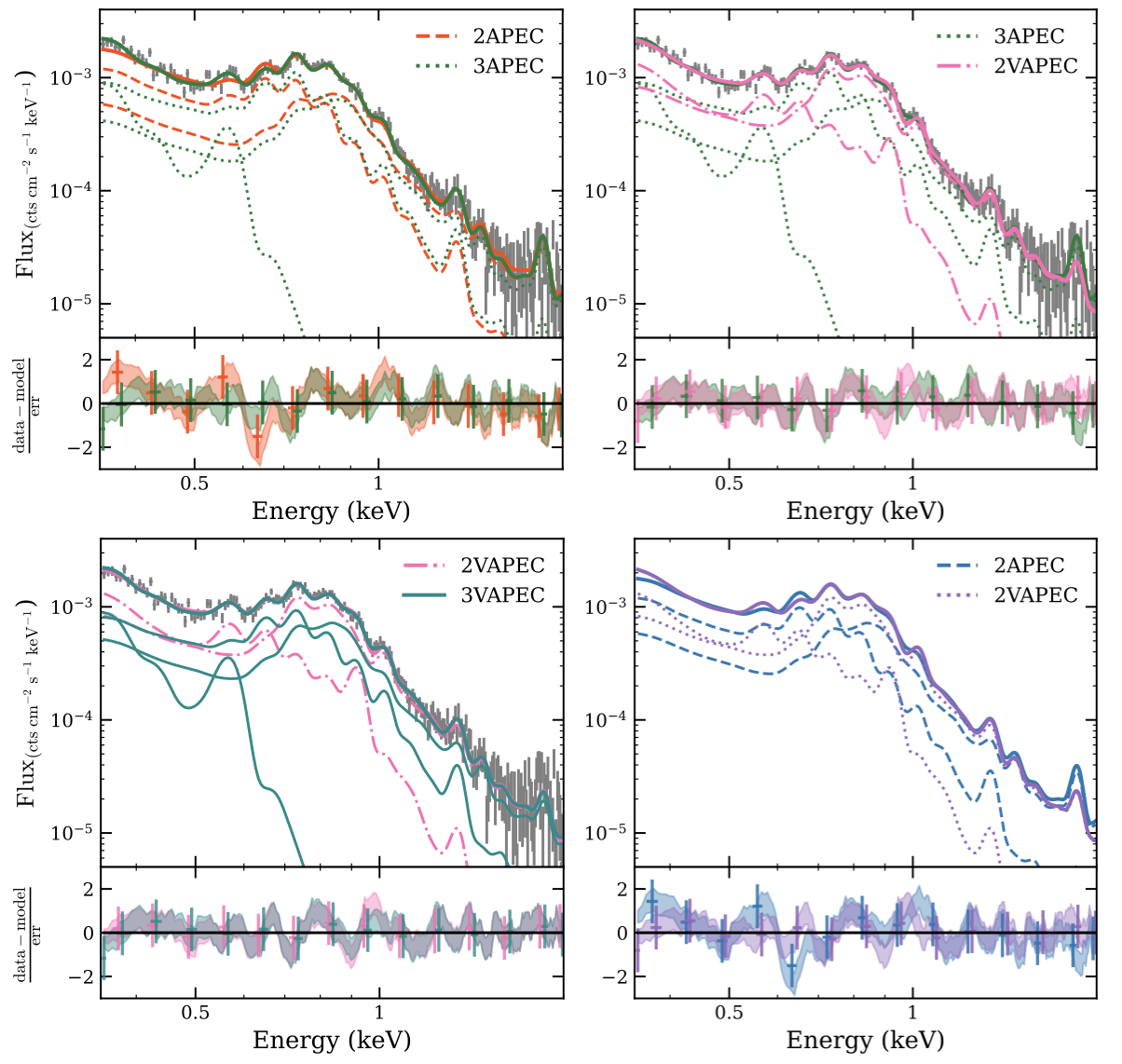}
    \caption{
Model fitting of the averaged X-ray spectra of 30 FGK stars (0.35--2.0~keV) using multi-temperature thermal plasma models. The top-left and top-right panels compare the performance of 2-temperature and 3-temperature APEC and VAPEC models, with corresponding residuals shown below each spectrum. The bottom-left panel illustrates the spectral fits from 2T-VAPEC and 3T-VAPEC models, highlighting differences in modelled line and continuum components. The bottom-right panel shows the comparison between APEC and VAPEC models, focusing on the impact of variable abundances on the fit. 
Residuals are expressed as (data - model)/error. 
The 2T-VAPEC model provides a notably satisfying fit under the level of d.o.f. in capturing both soft and hard X-ray features across the spectrum.}
    \label{fig:FGK_spec_fit_4model}
\end{figure*}

\subsection{Early-type M dwarfs vs. late-type M dwarfs}

The structural differences between early-type and late-type M dwarfs are particularly relevant for analysing their X-ray spectral properties: Early-type M dwarfs possess a radiative core, whereas mid/late-type M dwarfs are fully convective. This can influence their X-ray emission characteristics, which may be obscured if the two groups are combined in the analysis.

To investigate this distinction, we divide the 10 pc sample of M dwarfs into M0--M3.5 (early-type) and M4--M6.5 (mid/late-type) groups.
For clarity and to minimise potential biases, we exclude a pair of M stars, CD -37 10765 A (M3V) and B (M5V), which belong to different sub-spectra-type groups and are not distinguishable in X-ray. 
The final sample comprises 57 stars of subtype M0--M3.5 and 44 stars of subtype M4--M6.5, with two stars excluded.

Figure \ref{fig:M0346} presents the X-ray spectra of the two groups: the M0--M3.5 exhibit lower flux than M4--6.5 in the studied energy range, hence we scale up the flux of the M0--M3.5 group by a factor 1.73 to match the flux of the M4--M6.5 group at 0.7 keV for comparing. 
The top panel shows the fit using the two-temperature 2T-APEC model (see also in Table~\ref{tab:M0346}) and the middle panel displays the residuals of the fits. The bottom panel of Figure \ref{fig:M0346} shows the ratio between the scaled spectrum of the M0--M3.5 group and that of the M4--M6.5 group.

By visual inspection, the spectra reveal overall consistency in spectral shape despite the different luminosities, especially in the 0.5--2.0 keV range from the ratio plot. 
The mismatch rise in the very soft (<0.5 keV) band as the early-type M dwarfs display a relatively softer flux than the mid/late-type.

Table~\ref{tab:M0346} summarizes the spectral fits for M0--M3.5 and M4--M6.5 type M dwarfs using 3T-APEC models. Both fits achieve low reduced $\chi^2$.
The metal abundances are consistent between the two groups.
The temperatures of the intermediate and hot components (\(kT_1\) and \(kT_2\)) are in good agreement between the two groups. 
Interestingly, the M0--M3.5 stars in our sample are approximately 40\% fainter in X-ray luminosity compared to their M4--M6.5 counterparts, contrary to conventional expectations. 
Early-type M dwarfs, in which a radiative core remains beneath their convective envelopes, are generally thought to sustain higher coronal temperatures and produce more energetic flares, leading to stronger X-ray emission, particularly at harder energies. 
Late-type M dwarfs, on the other hand, are fully convective and often exhibit saturated magnetic activity. 
While this can maintain a high X-ray luminosity relative to their bolometric output, they are not usually expected to surpass earlier types in absolute luminosity due to $L_X-M_*$ dependence (see eg. \citealt{Magaudda2022AA}, \citealt{Magaudda2020AA}).
The observed trend suggests that additional effects, such as sample selection biases or flare activity, may contribute to the emission. 
It would therefore be valuable to test the luminosity–spectral-type relation with larger samples and assess whether the 10-pc stars are systematically different from common galactic population.

\begin{table*}[]
\centering
\caption{Model parameters and values for F, G, and K dwarf averaged spectrum.}
\begin{spacing}{1.4}
\centering
 \resizebox{\textwidth}{!}{
\begin{tabular}{ l | c c|c c |c c| c c}
\hline
\toprule
Param.	&	APEC	&	VAPEC	&	2T-APEC	&	2T-VAPEC	&	3T-APEC	&	3T-VAPEC	&		&	APEC+GAUSS\\
 \hline					

$kT_1$	&	${ 0.56} \pm 0.01$	&	${ 0.42^{+0.02}_{-0.01}}$	&	${ 0.31~(\pm 0.02, \pm0.02)}$	&	${ 0.22~(\pm0.01, {}^{+0.02}_{-0.01})}$	&	${ 0.09~(\pm0.01,\pm0.01)}$	&	$0.23~(\pm0.01,\pm0.01)$	&	$kT_1$	&	${ 0.51\pm0.01}$\\
$N_1$	&	${ 17.0 \pm 0.7}$	&	$24 \pm 2$	&	${ 8.0~(\pm0.9,{}^{+2.6}_{-0.1})}$	&	${ 10~(\pm 2, \pm2)}$	&	${ 10~({}^{+5}_{-3},{}^{+1}_{-3})}$	&	$9~{ (\pm3, {}^{+3}_{-2})}$	&	$N_1$	&	${ 17.4\pm0.9}$\\
$Z~or~A_{\rm Rest}$	&	${ 0.14} \pm 0.01$	&	${ 0.14\pm0.04}$	&	${ 0.31~(\pm 0.04, {}^{+0.06}_{-0.02})}$	&	$0.3~{ ({}^{+0.2}_{-0.1}, ^{+0.2}_{-0.1})}$	&	${ 0.55~({}^{+0.13}_{-0.09},{}^{+0.18}_{-0.06})}$	&	${ 0.4~({}^{+0.4}_{-0.2}, {}^{+0.4}_{-0.1})}$	&	$Z$	&	${ 0.13 \pm 0.01}$\\
$kT_2$	&		&		&	$0.75~(\pm 0.03, { {}^{+0.05}_{-0.01}})$	&	${ 0.60~(\pm0.02, \pm0.02)}$	&	$0.35~(\pm 0.02, { \pm0.01})$	&	${ 0.59~({}^{+0.06}_{-0.07},{}^{+0.04}_{-0.09})}$	&	$E_{G1}$	&	${ 0.57\pm0.01}$\\
$N_2$	&		&		&	${ 5.5~({}^{+0.8}_{-0.7},{}^{+1.0}_{-0.1})}$	&	${ 8~(\pm 2, {}^{+4}_{-1})}$	&	${ 4.6~(\pm0.8, {}^{+1.8}_{-0.1})}$	&	$5~{ (^{+3}_{-2}, {}^{+3}_{-2})}$	&	$N_{G1}$	&	${ 0.11\pm0.07}$\\
$kT_3$	&		&		&		&		&	$0.79~(\pm 0.03, { {}^{+0.04}_{-0.01})}$	&	${ 1.0~({}^{+0.1}_{-0.2}, {}^{+0.1}_{-0.3})}$	&	$E_{G2}$	&	${ 0.65~\mathrm{(fixed)}}$\\
$N_3$	&		&		&		&		&	${ 3.1~(\pm 0.6, {}^{+0.9}_{-0.3})}$	&	${ 1.5~({}^{+1.5}_{-0.9},{}^{+2.5}_{-0.8})}$	&	$N_{G2}$	&	${ 0.02\pm0.02}$\\
$A_\mathrm{O}$	&		&	${ 0.06} \pm 0.03$	&		&	${ 0.14~({}^{+0.06}_{-0.04},{}^{+0.08}_{-0.02})}$	&		&	${ 0.19~({}_{-0.07}^{+0.14}, {}^{+0.15}_{-0.06})}$	&	$E_{G3}$	&	${ 0.915^{+0.009}_{-0.007}}$\\
$A_\mathrm{Ne}$	&		&	${ 0.44^{+0.06}_{-0.05}}$	&		&	${ 0.8~({}^{+0.3}_{-0.2}, {}^{+0.4}_{-0.1})}$	&		&	${ 0.7~({}^{+0.4}_{-0.3}, {}^{+0.5}_{-0.2})}$	&	$N_{G3}$	&	${ 0.35 \pm 0.05}$\\
$A_\mathrm{Si}$	&		&	${ 0.22^{+0.09}_{-0.08}}$	&		&	${ 0.18~({}^{+0.12}_{-0.09}, {}^{+0.01}_{-0.18})}$	&		&	${ 0.2~({}^{+0.2}_{-0.1}, {}^{+0.1}_{-0.2})}$	&	$E_{G4}$	&	${ 1.01\pm0.01}$\\
$A_\mathrm{Fe}$	&		&$0.12^{+0.02}_{-0.01}$	&		&	${ 0.27~({}^{+0.09}_{-0.06}, {}^{+0.08}_{-0.06})}$	&		&	${ 0.4~({}^{+0.3}_{-0.1}, {}^{+0.3}_{-0.1})}$	&	$N_{G4}$	&	${ 0.14} \pm 0.03$\\
 \hline						
$\chi_{\nu}^2\ (\frac{\chi^2}{\mathrm{d.o.f.}})$	&	${ 2.58~(\frac{574.48}{223})}$	&	$1.94~(\frac{424.11}{219})$	&	${ 1.58~(\frac{349.13}{221})}$	&	${ 1.08~(\frac{235.16}{217})}$	&	${ 1.15~(\frac{252.77}{219})}$	&	${ 1.07~(\frac{229.39}{215})}$	&	$\chi_{\nu}^2\ (\frac{\chi^2}{\mathrm{d.o.f.}})$	&	${ 1.49~(\frac{321.25}{216})}$\\
AIC	&	{ 580.48}	&	438.11 &	{ 359.13}	&	{ 253.16}	&	{ 266.77}	&	{ 251.39}	&	AIC	&	{ 341.25}\\
BIC	&	{ 590.74}	&	{ 436.53}	&	{ 376.23}	&	{ 283.94}	&	{ 290.71}	&	{ 289.01}	&	BIC	&	{ 375.46}\\
\bottomrule
\end{tabular}}
\end{spacing}
\tablefoot{Fit is conducted in energy range of 0.35--2.0 keV. Normalization $N_{\rm apec}$ is in the unit $10^{-4}\frac{{10}^{-14}}{4\pi(10 {\rm pc})^2}~ \mathrm{cm}^{-5}$. The set of abundance is bonded for all components of a models. The error reported is corresponds to 90\% confidence, assuming chi-square statistics. In model 2T-VAPEC, 3T-APEC, and 3T-VAPEC, the second error is the 90\% confidence interval from the bootstrap resampling distribution (Section~\ref{sec:bootstrap}). $A_{\rm element}=\frac{(n_{Z}/n_{H})_{\rm plasma}}{(n_{Z}/n_{H})_{\odot}}$.}\label{tab:FGKfit}
\end{table*}

\begin{table}[h]
\caption{F-test statistics of models for average FGK star.}
    \centering    
    \resizebox{\columnwidth}{!}{ 
    \begin{tabular}{cc|cc|cc}
        \hline
        \toprule
Simpler	&	More~complex	&	\multirow{2}{*}{$\Delta$ d.o.f.}	&	\multirow{2}{*}{$\Delta \chi^2$}	&	\multirow{2}{*}{F-value}	&	\multirow{2}{*}{p-value}	\\
Model & Model & & & & \\
\hline
1T-APEC	&	2T-APEC	&	2	&	225.35	&	71.3	&	$1\times10^{ -24}$	\\
2T-APEC	&	3T-APEC	&	2	&   96.36	&	41.7	&	$4\times10^{ -16}$	\\
\hline													
1T-APEC	&	1T-VAPEC	&	4	&	{ 150.37}	&	{ 19.4}	&	${ 1\times10^{-13}}$	\\
2T-APEC	&	1T-VAPEC	&	2	&	{ -74.98}	&	<0	&	-	\\
2T-APEC	&	2T-VAPEC	&	4	&	{ 113.97}	&	{ 26.3}	&	${ 9\times10^{-18}}$	\\
3T-APEC	&	2T-VAPEC	&	2	&	{ 17.61}	&	{ 8.1}	&	${ 4\times10^{-4}}$	\\
3T-APEC	&	3T-VAPEC	&	4	&	{ 23.38}	&	{ 5.5}	&	${ 3\times10^{-4}}$	\\
\hline													
1T-VAPEC	&	2T-VAPEC	&	2	&	{ 188.95}	&	{ 87.2}	&	${ 2\times10^{-28}}$	\\
2T-VAPEC	&	3T-VAPEC	&	2	&	{ 5.77}	&	{ 2.7}	&	${ 0.069}$	\\
     \bottomrule
    \end{tabular}}
    \tablefoot{The improvement in best-fit between two models regarding change in degrees of freedom ($\Delta$ d.o.f.), change in $\chi^2$, the corresponding F-value, and the resulting p-value. The d.o.f. is $N_\mathrm{data}-N_\mathrm{free~param.}$.}
    \label{tab:model_ftest_comparison_fgk}
\end{table}

\section{Results: F, G, K-type stars sample}
\label{sec:fittingFGK}

In this section, we apply the same spectral fitting approach used for the M dwarfs to the stacked spectra of FGK-type stars within the 10~pc sample. 
The spectrum shown in Figure~\ref{fig:FGK_spec_fit_4model} is constructed from 30 FGK-type dwarf stars, all included in the \erosita\_DE sky survey. 
While this dataset comprehensively measures nearby FGK-type stars, the limited sample size may not fully capture the diversity of coronal properties present within this stellar class.
FGK dwarfs are brighter than M dwarfs in the optical and ultraviolet bands. A high optical and UV brightness potentially introduces optical loading on the SRG/eROSITA detectors. This effect alters the information on the detected photons (number, energy, and detection pattern) and becomes increasingly important at lower energies. In fact, a significant number of these FGK stars are potentially affected by optical loading. To retain all 30 FGK-type stars in the analysis, we restrict our spectral fitting to energies above 0.35 keV. By doing so, we mitigate the influence of optical loading, whose effect is only significant at lower energies (see Appendix\ref{sec:opt_appdix}). 
We note that this treatment largely helps reduce the flux excess at the soft end; however, it is important to note that the “blue-shift” caused by optical loading affects the entire energy range. 
{ We present spectra affected by and free from optical loading below 0.35 keV in Appendix~\ref{sec:opt_appdix} and test them with 3T model.}

The parameters obtained from all fitted models are summarized in Table~\ref{tab:FGKfit}. 
Similarly to the M dwarf case, the fit using a single-temperature APEC model (1T-APEC) yields statistically unacceptable results, with a reduced chi-squared of $\chi^2_\nu = { 2.58}$. The same considerations are valid on adding complexity to this simple model. We thus do not discuss this model further. 
The 1T-VAPEC model, which allows metal abundances to vary, also provides a poor fit with $\chi^2_\nu = 1.94$. Together with the inadequate fit of the 1T-APEC model, this result suggests that a single-temperature description is insufficient—even when elemental abundances are allowed to vary.

Therefore, we added a second APEC model (2T-APEC; top-left panel in Figure~\ref{fig:FGK_spec_fit_4model}), which significantly improved the fit, reducing the chi-squared by $\Delta\chi^2 = { 225.35}$ for two additional degrees of freedom (see Table~\ref{tab:FGKfit}).

The 2T-VAPEC model further improves the fit compared to both the 1T-VAPEC model and the 2T-APEC model in regard of F-test (see Table~\ref{tab:model_ftest_comparison_fgk}). 
The best-fit temperatures for the two components are ${ 0.22\pm0.01}$ keV and ${ 0.60\pm0.02}$ keV, with corresponding abundances of $A_{\ion{O}{}} = { 0.14^{+0.06}_{-0.04}},~A_{\ion{Ne}{}} = { 0.8^{+0.3}_{-0.2}}$, $A_{\ion{Si}{}} = { 0.18^{+0.12}_{-0.09}}$, $A_{\ion{Fe}{}}= { 0.27^{+0.09}_{-0.06}}$ and a shared abundance of ${ 0.3^{+0.2}_{-0.1}}$ for the rest elements.

Instead of letting the metal abundances free to vary, another way to add complexity with respect to the 2T-APEC model is to introduce a third thermal component (with linked abundance across all components). This 3T-APEC fit ($\chi^2_\nu = 1.15$) also shows improvement relative to the 2T-APEC model.
{ The F-test comparison, $F(2, 96.36)$, also supports the 3T-APEC model over the 2T-APEC configurations. 
The best-fit temperatures from 3T-APEC model are $kT_1 = 0.09\pm0.01$ keV, $kT_2 = 0.35\pm0.02$ keV, and $kT_3 = 0.79\pm0.03$ keV, with a shared abundance of $0.55^{+0.13}_{-0.09}$.
The presence of the very soft component ($kT_1 = 0.09\pm0.01$) could be a residual effect of optical loading. }

\begin{table*}[]
\centering
\begin{minipage}{0.47\linewidth}
\centering
\caption{Fluxes and Luminosities of Stacked M-star Spectra}
\label{tab:eneflux}
\begin{spacing}{1.2}
\resizebox{\linewidth}{!}{
\begin{tabular}{l c|cc}
\hline
\toprule
Model	&	Energy range	&	Flux@10 pc	&		Luminosity	\\
	&	 \tiny{(keV)}	&	\tiny{$(\mathrm{10^{-13}~erg/cm\textsuperscript{2}/s})$}	&	\tiny{$(\mathrm{10^{27}~erg/s})$}\\
\hline									
\multirow{4}{*}{2T-APEC}	&	0.2 -- 2.0	&	$2.2\pm0.1$	&	$2.6\pm0.1$	\\
	&	0.2 -- 0.5	&	$0.74\pm0.04$	&	$0.9\pm0.1$	\\
	&	0.5 -- 1.0	&	$1.06\pm0.03$	&	$1.27\pm0.04$	\\
	&	1.0 -- 2.0	&	$0.39\pm0.02$	&	$0.46\pm0.02$	\\
        &   0.1 -- 2.4  &   $2.7\pm0.1$  &   $3.2\pm0.1$ \\ 
    \hline
\multirow{2}{*}{$T_1$(0.27 keV)}	&	0.2 -- 2.0	&	$1.08\pm0.08$	&	$1.3\pm0.1$	\\
     &   0.1 -- 2.4  &   $1.4\pm0.1$        &   $1.7\pm0.1$ \\
    \hline
\multirow{2}{*}{$T_2$(0.94 keV)}	&	0.2 -- 2.0	&	$1.11\pm0.07$	&	$1.33\pm0.08$	\\
    &   0.1 -- 2.4  &   $1.33\pm0.08$        &   $1.6\pm0.1$ \\
\hline
\bottomrule
\multirow{4}{*}{2T-VAPEC}	&	0.2 -- 2.0	&	$2.2\pm0.1$	&	$2.6\pm0.1$	\\
	&	0.2 -- 0.5	&	$0.76\pm0.08$	&	$0.9\pm0.1$	\\
	&	0.5 -- 1.0	&	$1.06\pm0.05$	&	$1.26\pm0.05$	\\
	&	1.0 -- 2.0	&	$0.38\pm0.03$	&	$0.46\pm0.03$	\\
        &   0.1 -- 2.4  &   $2.7\pm0.1$        &   $3.3\pm0.2$ \\
    \hline
\multirow{2}{*}{$T_2$(0.26 keV)}	&	0.2 -- 2.0	&	$1.11\pm0.08$	&	$1.3\pm0.1$	\\
        &   0.1 -- 2.4  &   $1.4\pm0.1$        &   $1.7\pm0.1$ \\
    \hline
\multirow{2}{*}{$T_3$(0.96 keV)}	&	0.2 -- 2.0	&	$1.09\pm0.04$	&	$1.3\pm0.05$	\\
    &   0.1 -- 2.4  &   $1.31\pm0.05$        &   $1.57\pm0.05$ \\
\bottomrule
\end{tabular}}
\end{spacing}
\end{minipage}\hfill
\begin{minipage}{0.48\linewidth}
\centering
\caption{Fluxes and Luminosities of Stacked FGK-star Spectra.}
\label{tab:eneflux_fgk}
\begin{spacing}{1.2}
\resizebox{\linewidth}{!}{%
\begin{tabular}{lc|cc}
\hline
\toprule
Model	&	Energy Range	&	Flux@10 pc	&		Luminosity	\\
	&	 \tiny{(keV)}	&	\tiny{$(\mathrm{10^{-13}~erg/cm\textsuperscript{2}/s})$}	&	\tiny{$(\mathrm{10^{27}~erg/s})$}\\
\hline									
\multirow{4}{*}{2T-VAPEC}	&	0.2 -- 2.0	&	$13\pm2$	&	$15\pm3$	\\
	&	0.2 -- 0.5	&	$4.5\pm0.9$	&	$5\pm1$	\\
	&	0.5 -- 1.0	&	$6\pm2$	&   $8\pm2$	\\
	&	1.0 -- 2.0	&	$1.8\pm0.9$	&	$2.2\pm0.8$	\\
    &   0.1 -- 2.4  &   $16\pm4$  &   $19\pm4$  \\
    \hline
\multirow{2}{*}{$T_1$(0.22 keV)}	&	0.2 -- 2.0	&	$5\pm1$	&	$6\pm1$	\\
	&	0.1 -- 2.4	&	$7\pm1$	&	$8\pm1$	\\
    \hline
\multirow{2}{*}{$T_2$(0.60 keV)}	&	0.2 -- 2.0	&	$8\pm2$	&	$9\pm2$	\\
	&	0.1 -- 2.4	&	$9\pm1$	&	$11\pm2$	\\
\hline
\bottomrule
\multirow{4}{*}{3T-VAPEC}	&	0.2 -- 2.0	&	$13\pm2$	&	$15\pm3$	\\
	&	0.2 -- 0.5	&	$4.4\pm0.8$	&	$5\pm1$	\\
	&	0.5 -- 1.0	&	$7\pm1$	&   $8\pm2$	\\
	&	1.0 -- 2.0	&	$1.8\pm0.4$	&	$2.1\pm0.6$	\\
    &   0.1 -- 2.4  &   $16\pm3$  &   $19\pm4$  \\
    \hline
\multirow{2}{*}{$T_1$(0.23 keV)}	&	0.2 -- 2.0	&	$5.6\pm0.7$	&	$6.7\pm0.8$	\\
	&	0.1--2.4	&	$7.4\pm0.8$	&	$8.8\pm0.9$	\\
    \hline
\multirow{2}{*}{$T_2$(0.59 keV)}	&	0.2 -- 2.0	&	$6\pm1$	&	$7\pm2$	\\
	&	0.1 -- 2.4	&	$6\pm2$	&	$8\pm2$	\\
    \hline
\multirow{2}{*}{$T_3$(1.0 keV)}	&	0.2 -- 2.0	&	$1.6\pm0.5$	&	$2.0\pm0.6$	\\
	&	0.1 -- 2.4	&	$1.9\pm0.6$	&	$2.3\pm0.7$	\\
\bottomrule
\end{tabular}}
\end{spacing}
\end{minipage}
\tablefoot{The fluxes and luminosities are derived from the best-fit models of fitting spectra in 0.2--2.0 keV for M stars and 0.35--2.0 keV for FGK stars. For the 0.1--2.4 keV band, corresponding to the ROSAT band, the values are obtained by extrapolating these best-fit models. Fluxes are presented as if one average source were located at 10 pc. The uncertainties here are the conservative union of the 90\% $\chi^2$ statistical confidence intervals and the systematic offset between the AC and AR stacking methods.}
\end{table*}

{ An F-test comparing the 2T-VAPEC and 3T-APEC fits yields (F(2,17.61)=8.1), exceeding the 3$\sigma$ threshold and favouring 2T-VAPEC over 3T-APEC. This indicates that improving the 2T-APEC description by allowing metal abundances to vary is more effective than adding an extra thermal component. At the same time, while 3T-VAPEC improves upon 3T-APEC, it does not outperform 2T-VAPEC: the comparison between 3T-VAPEC and 2T-VAPEC gives (F(2,5.77)=2.7) (p = 0.069), below 3$\sigma$ significance. Considering $\Delta$AIC and $\Delta$BIC alongside the F-tests, we adopt 2T-VAPEC as the conservative model for the subsequent discussion. }

Although this analysis includes all FGK stars within the 10 pc WGH, the limited sample size may not provide a sufficiently comprehensive spectroscopic view of the unified properties of local FGK stars. 
Further studies employing higher-resolution spectra and/or larger samples will be essential to refine these preliminary findings, especially regarding the softest end of the energy band considered here.

\begin{figure}
    \centering
    \includegraphics[trim={0.3cm 0.2cm 0.3cm 0.1cm}, clip, width=\linewidth]{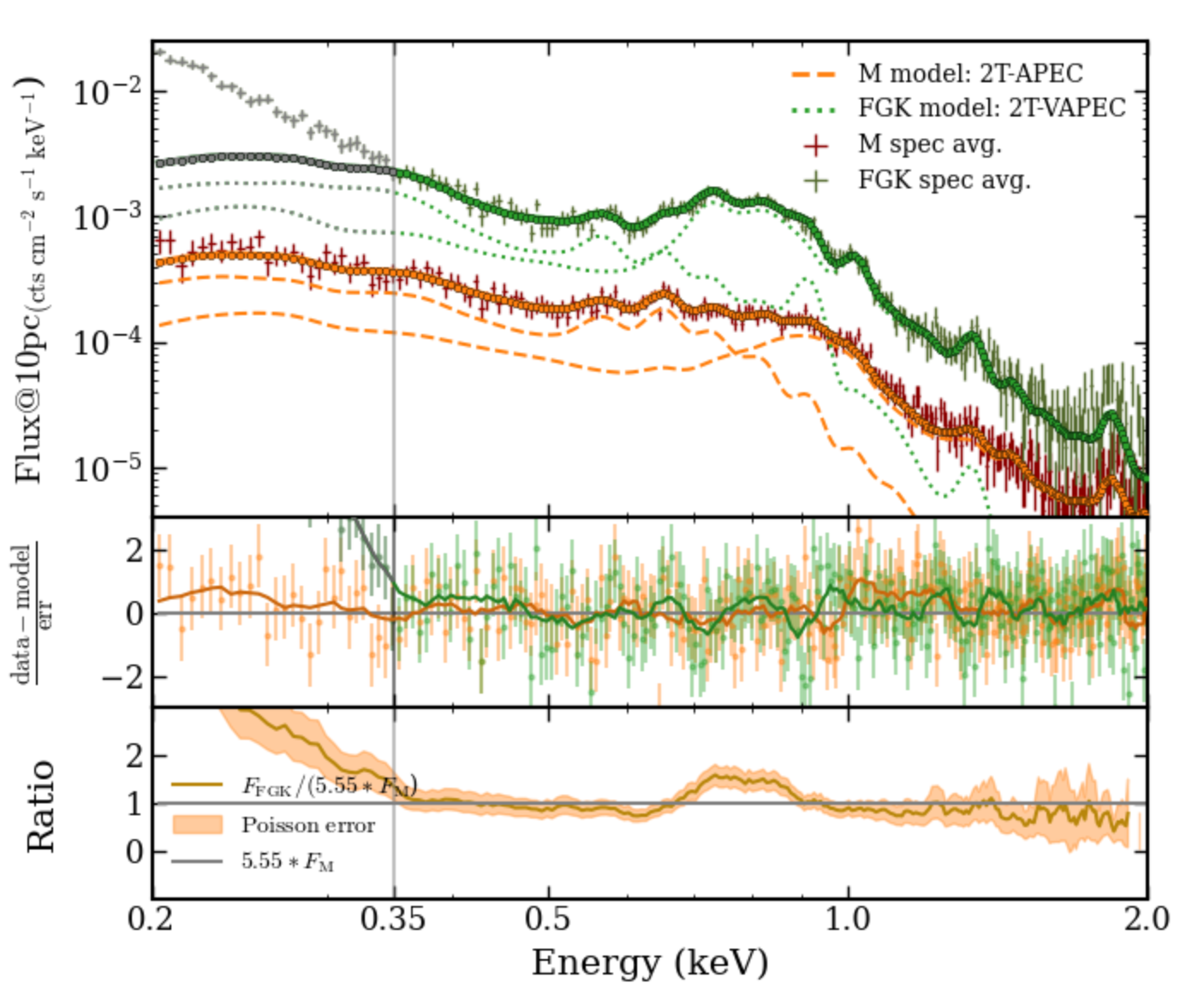}
    \caption{Comparison of the spectra of 103 M stars (orange) and 30 FGK stars (green) in 10pc-\gaia\ sample. Model 2T-apec and 2T-VAPEC are applied respectively.
    The middle panel displays the residual of fit.
    The bottom panel shows the ratio between the average FGK and 5.55 times brighter M-star spectra. The scale comes from the ratio of luminosity in 0.2-2.0 keV. The Poisson noise is applied.}
    \label{fig:summary}
\end{figure}

\section{Results: average luminosities}

In Table~\ref{tab:eneflux}, we report the flux and luminosity in different energy bands for the eROSITA\_DE stacked 10 pc { (WGH)} M dwarf spectrum based on the 2T-APEC and { 2T-VAPEC} models.
The flux from component 0.27 keV and 0.94 keV is reported in 0.2--2.0 keV and extrapolated for 0.1--2.4 keV, which is the band pass of the ROSAT PSPC.
We also derive FGK average luminosities for the 2T-VAPEC and 3T-VAPEC models in Table~\ref{tab:eneflux_fgk}.
{ We define and report conservative uncertainties for flux and luminosity by accounting for the differences between the AC and AR stacking methods.
Specifically, the quoted uncertainties are taken as the envelope (union) of the 90\% $\chi^2$ statistical confidence and intervals and the systematic offset between results produced by AC and AR.
This provides a lower bound on the total uncertainty compared to that derived from model fitting. }

In the 0.2--2.0 keV band, M dwarfs exhibit an average luminosity of $(2.6\pm0.1)\times10^{27}$ erg s$^{-1}$ (2T-APEC), while FGK stars have $(15\pm3)\times10^{27}$ erg s$^{-1}$ (2T-VAPEC). Thus, the averaged FGK star in 10 pc (WGH) is more than 5 times brighter than an M dwarf in this energy range.
In Figure~\ref{fig:summary}, we present the spectra of M stars and FGK stars overlaid with their respective model fits. 
By comparing the M-star and FGK-star spectra in the bottom panel of Figure~\ref{fig:summary}, we find the overall similarity except the primary difference arising in the 0.7--0.9 keV energy range. In Figure~\ref{fig:separate_erass}, we also show there is no significant systematic variation in the average luminosity is observed across the four eRASS epochs.

Independent of this 10-WGH sample considered in this work, early-type M stars (M0–M4) in the EGH appear to host more high luminosity sources than the WGH. Across the combined WGH+EGH 10pc-\gaia\ sample, M dwarfs can exhibit a mean luminosity of up to $\sim6\times10^{27}$ erg s$^{-1}$, as inferred from \citet{Caramazza2023AA}. By comparison, the FGK sample shows no measurable hemispheric dependence in its average luminosity.

\section{Conclusions}\label{sec:conclu}
In this study, we provide the highest signal-to-noise distance-normalized spectrum of M dwarfs and FGK-type stars by stacking over the volume-complete sample of stars within 10~pc in the western Galactic hemisphere, using data from the first four X-ray surveys of the sky performed by the SRG/\erosita\ instrument (eRASS1 to 4). 
The average X-ray luminosity of WGH in 0.2--2.0 keV is of $\rm (2.6 \pm0.1)\times 10^{27} ~erg/s$ for 10 pc M-type stars, and $\rm (15\pm3)\times 10^{27}~erg/s$ for F, G, and K-type stars.
When considering the entire 10pc-\gaia\ sample in both WGH and EGH, the average luminosity for the M star sample could be as high as $\sim 6\times10^{27}~\mathrm{erg/s}$, cross-estimated from \citet{Caramazza2023AA}, whereas the average luminosity of FGK stars shows no significant difference between the two hemispheres.
The luminosities derived here are in good agreement across all four eRASS surveys.
The stacked spectra are well reproduced by collisionally ionized plasma components at different temperature.
The fitted temperature structure and abundances remain consistent despite the presence of the most luminous outliers in the brightness distribution.
Specifically, all M-star subsamples tested exhibit the same two thermal components, at approximately 0.27 keV and 0.94 keV.
Additional softer or harder components are required depending on the energy range considered and on-going flaring activity.
Apart from the emission measure, no evidence of systematic differences in the fitted parameters are found between the stacked spectra of early-type (M0--M3.5) and mid-to-late-type (M4--M6.5) M dwarfs. Interestingly, the early-M stars appear on average less luminous than the mid/late-M stars, in contrast to the expected $L_\mathrm{X}$–$M_{*}$ trend \citep{Magaudda2022AA}.

Our study provides insights into nearby low-mass stars' X-ray luminosity properties.  
In this work we used the highest statistics from the combined four \erosita\ All-Sky Surveys. Potential improvement may come from the availability of models to account for optical loading.

\begin{acknowledgements}
This work is based on data from eROSITA, the soft X-ray instrument aboard SRG, a joint Russian-German science mission supported by the Russian Space Agency (Roskosmos), in the interests of the Russian Academy of Sciences represented by its Space Research Institute (IKI), and the Deutsches Zentrum für Luft und Raumfahrt (DLR). The SRG spacecraft was built by Lavochkin Association (NPOL) and its subcontractors, and is operated by NPOL with support from the Max Planck Institute for Extraterrestrial Physics (MPE).

The development and construction of the eROSITA X-ray instrument was led by MPE, with contributions from the Dr. Karl Remeis Observatory Bamberg \& ECAP (FAU Erlangen-Nuernberg), the University of Hamburg Observatory, the Leibniz Institute for Astrophysics Potsdam (AIP), and the Institute for Astronomy and Astrophysics of the University of Tübingen, with the support of DLR and the Max Planck Society. The Argelander Institute for Astronomy of the University of Bonn and the Ludwig Maximilians Universität Munich also participated in the science preparation for eROSITA.
    The eROSITA data shown here were processed using the eSASS/NRTA software system developed by the German eROSITA consortium.

    We acknowledge financial support from the European Research Council (ERC) under the European Union’s Horizon 2020 research and innovation program Hot-Milk (grant agreement No 865637). GP acknowledges support from Bando per il Finanziamento della Ricerca Fondamentale 2022 dell’Istituto Nazionale di Astrofisica (INAF): GO Large program and from the Framework per l’Attrazione e il Rafforzamento delle Eccellenze (FARE) per la ricerca in Italia (R20L5S39T9). MF and MY acknowledge support from the Deutsche Forschungsgemeinschaft through the grant FR 1691/2-1.
\end{acknowledgements}

\bibliography{aa} 

\begin{appendix}

\section{The impact of optical loading on the spectrum of bright stars}\label{sec:opt_appdix}

When an X--ray photon reaches the depletion layer of an X--ray CCD, it creates electron--hole pairs there \citep{Dennerl2020SPIE, Meidinger2021JATISCCD}. The same may happen when an optical photon reaches this layer. In contrast to an X--ray photon, however, where the number of released electrons can be utilized for spectroscopy, the absorption of a single optical photon releases typically only one electron. This charge alone is not high enough to get included into the telemetry stream, because it is below the low energy threshold, which needs to be applied in order to prevent telemetry overload due to electronic noise. In the case of celestial sources which are both optically and X--ray bright, however, the similar (though not identical) optical and X-ray point-spread function (PSF) favour the case that the charge clouds created by the optical and X-ray photons get superimposed within the integration time of a particular CCD frame. In the following we discuss the consequences of that case.

\medskip\noindent
For better clarity we start with a simplified example, which considers the presence of only one X-ray and one optical photon and assumes absence of electronic noise. In that case, the observed spectrum is modified due to the additional charge generated by the optical photon in two ways:

\vspace*{-1mm}
\begin{itemize}

\item{ energy:}
the spectrum gets `blue-shifted', and because of the Poissonian nature of the additional charge, it gets (somewhat) smeared; while these effects are most apparent at low energies, they affect the whole spectrum.

\smallskip
\item{ flux:}
the geometry and extent of the charge cloud get modified, causing a change of the reconstructed flux.

\end{itemize}

\noindent
While the consequences of optical loading on the energy is straightforward, its impact on the flux requires a more detailed investigation. We need to consider that the charge cloud created by an X--ray photon may extend over up to four pixels, which we refer to as \texttt{`singles (s)'}, \texttt{`doubles (d)'}, \texttt{`triples (t)'}, and \texttt{`quadruples (q)'}. The \texttt{triples} and \texttt{quadruples} have to meet the following criteria in order to have been generated by an X--ray photon: the \texttt{triples} must be L-shaped, with the main pixel (the one containing the dominant charge component) at the corner, and the  quadruples must be quadratic, with the main pixel essentially opposite to the pixel containing the minimum charge. Only such \texttt{triples} and \texttt{quadruples} are considered as `valid' and are used for reconstructing the flux (for the \texttt{singles} and \texttt{doubles} there is no restriction on their validity other than that they must be sufficiently far away from an insensitive area). The additional charge generated by an optical photon can then modify the pixel pattern in several ways:

\vspace*{-1mm}
\begin{enumerate}
	
\item{ it neither changes the pattern size nor the validity of the pattern:} then only the energy is affected, but not the flux (when taking the energy shift into account); this holds also for the pattern-specific flux, derived, e.g., from considering only singles.

\item{ it increases the pattern size, but keeps the pattern valid:} then the energy and the pattern-specific flux are affected, but not the `sdtq' flux; the increase of the pattern size is caused by raising the charge in a neighbouring pixel above the low energy threshold.

\item{ it makes the pattern invalid:} then the X--ray photon gets lost and the flux is reduced, but the energy scale is not affected.

\end{enumerate}

\noindent
If we consider the presence of several optical photons (optical pile-up), then this situation stays unaffected as long as the optical photon flux is not high enough to generate charge above the low energy threshold without an X--ray photon. However, when this happens, the affected pixels have to be masked out in order to avoid saturation of telemetry. Therefore, this case does not need further consideration. Similarly, the presence of several X--ray photons (X--ray pile-up) as well as the combination of optical and X--ray pile--up does not need to be considered here, because it might prevent a spectroscopic analysis.

\smallskip\noindent
What still needs to be considered is the presence of electronic noise, which is steeply rising towards the low energy threshold (which is the reason why such a threshold needs to be applied). Superposition with the charge released by optical photons causes its spectrum to get `blue--shifted' in a similar way as an X--ray spectrum, making it appear amplified and creating the impression of the presence of an additional soft `optical' component. While to some extent this effect is mitigated by considering the local background in the spectral analysis, an apparent soft energy component may remain. For completeness we mention that there is in addition particle induced background present, which may also be affected by optical loading. This component, however, is comparatively faint.

\subsection{Between patterns and TMs}

The effects described above can be utilized for studying the eROSITA spectra for evidence of optical loading. In the following we present FGK spectra which are divided by the effective area (indicated by `cm$^{-2}$') for better clarity as significant number of FGK stars are optically bright, with G-band magnitudes below 5, potentially causing strong optical loading.

\begin{figure*}[]
  \centering
 \begin{subfigure}{0.32\textwidth}
 \includegraphics[width=\linewidth]{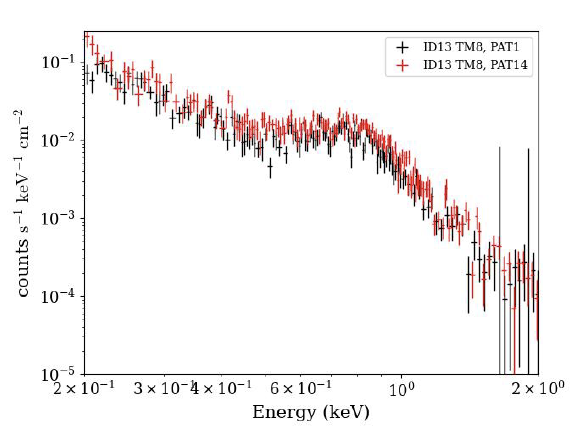}
\caption{TM8 spectra for source \#13: `s' (black) and `dtq' (red).\label{fig01}}
\end{subfigure}
\begin{subfigure}{0.32\textwidth}
 \includegraphics[width=\linewidth]{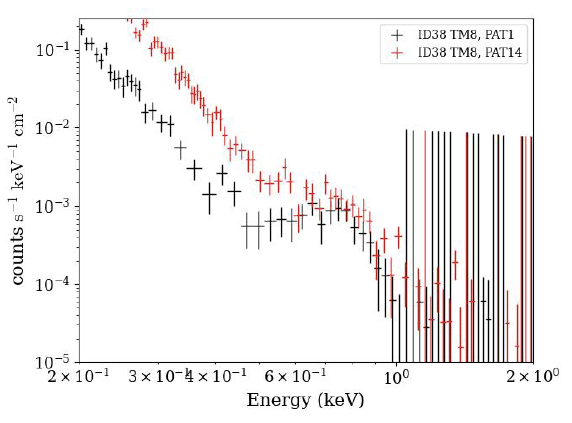}
\caption{TM8 spectra for source \#38: `s' (black) and `dtq' (red).\label{fig02}}
\end{subfigure}
  \begin{subfigure}{0.32\textwidth}
    \includegraphics[width=\linewidth]{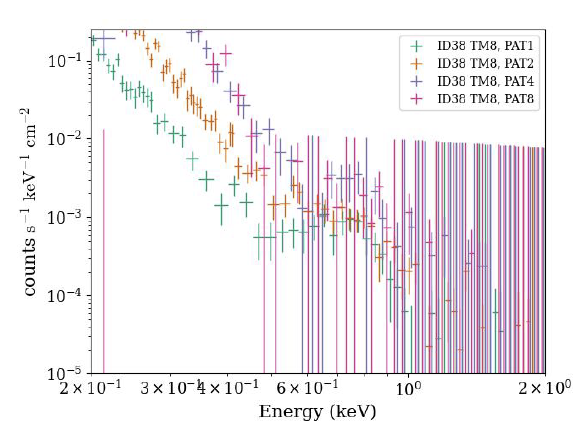}
    \caption{TM8 spectra for source \#38: `s' (green), `d' (blue), `t' (purple), and `q' (red).\label{fig03}}
  \end{subfigure}

  \begin{subfigure}{0.32\textwidth}
    \includegraphics[width=\linewidth]{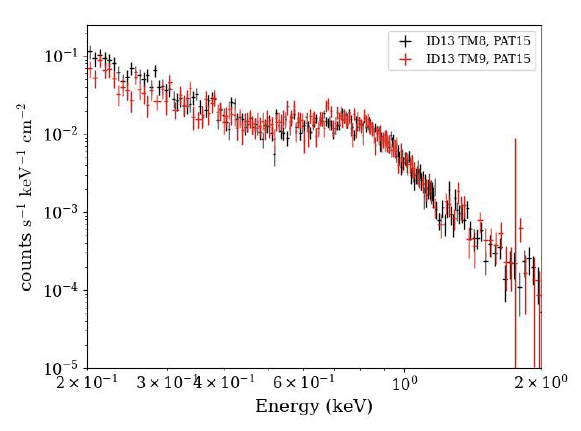}
    \caption{`sdtq' spectra for source \#13: TM8 (black) and TM9 (red).\label{fig08}}
  \end{subfigure}
  \begin{subfigure}{0.32\textwidth}
    \includegraphics[width=\linewidth]{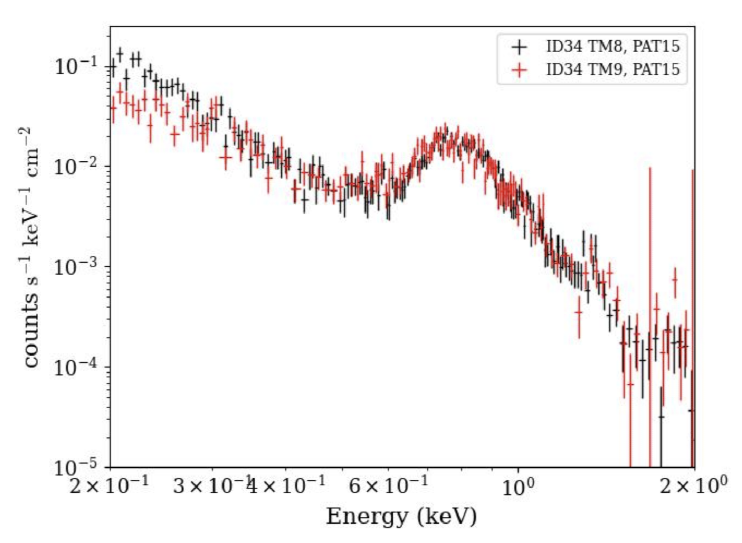}
    \caption{`sdtq' spectra for source \#34: TM8 (black) and TM9 (red).\label{fig04}}
  \end{subfigure}
  \begin{subfigure}{0.32\textwidth}
    \includegraphics[width=\linewidth]{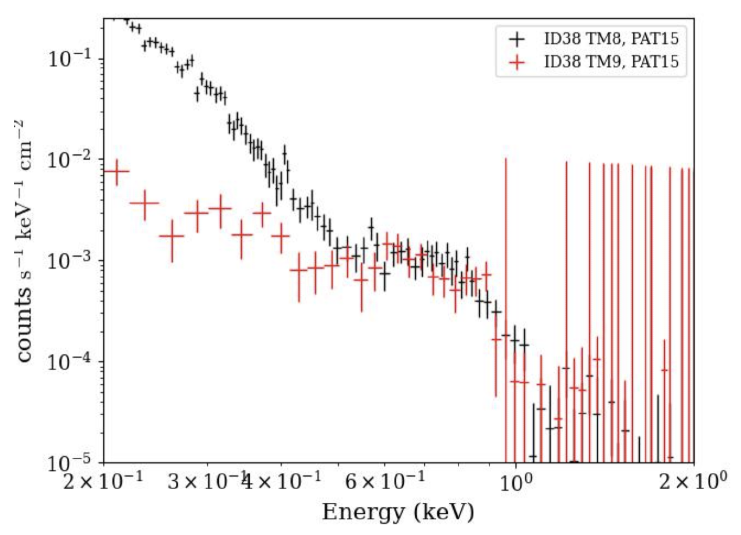}
    \caption{`sdtq' spectra for source \#38: TM8 (black) and TM9 (red).\label{fig05}}
  \end{subfigure}

  \caption{Event-pattern and -TM selection effects for three optically bright stars: eps Eri (\#13, K2V star with G3.5~mag), $\pi^3$~Ori (\#34; F6V star with G3.2~mag), and $\beta$ Hy (\#38; G0V subgiant with G2.7~mag).}
  \label{fig:sixplots}
\end{figure*}

\smallskip\noindent
Figure \ref{fig01} investigates the effect 1 by comparing pattern-specific spectra: \texttt{`s'} (black) with \texttt{`dtq'} (red). The fact that \texttt{`s'} is only slightly below \texttt{`dtq'} indicates that the pattern size and validity of the patterns are not much affected, indicating only minor optical loading. In contrast, Figure\,\ref{fig02} presents a case where \texttt{`s'} is considerably below \texttt{`dtq'}, which is evidence for substantial optical loading. Subdividing the \texttt{`dtq'} spectrum of Figure\,\ref{fig02} into the individual \texttt{`d'}, \texttt{`t'}, and \texttt{`q'} components (Figure\,\ref{fig03}) shows that the apparent spectral flux increases with pattern size up to triples. This is consistent with effect 2: if the charge cloud created by the absorption of an X--ray photon is distributed over more pixels, then the probability that an optical photon hits a pixel in that cloud is increased. This, however, does not apply to quadruples, because any extension beyond four pixels makes a pattern invalid, leading to a loss of flux (effect 3). This loss, however, is quite small, because the probability that quadruples are created by an X--ray photon below 0.5~keV is less than 3\%. 

\smallskip\noindent
Another way of checking for optical loading is to compare the spectra between \texttt{TM8} and \texttt{TM9}, as they differ in the thickness of the Al layer in the optical blocking filter, which is 200 nm for \texttt{TM8} and 100 nm for \texttt{TM9} \citep{Meidinger2021JATISCCD}. Thus, \texttt{TM9} is more sensitive to optical flux than \texttt{TM8}. Figure~\ref{fig08} shows that for source \#13 (the same as in Figure\,\ref{fig01}) the \texttt{`sdtq'} spectra for \texttt{TM8} and \texttt{TM9} (when divided by the effective area) are similar. This indicates that the derived luminosity is not much affected by optical loading (effect 2), consistent with the conclusion above, which was based on a comparison between the \texttt{`s'} and \texttt{`dtq'} spectra for \texttt{TM8} (effect 1, Figure\,\ref{fig01}). For source \#34 (Figure\,\ref{fig04}), however, \texttt{TM9} is clearly below \texttt{TM8}, and for source \#38 (Figure\,\ref{fig05}), \texttt{TM9} is considerably below \texttt{TM8} a clear indication for optical loading. Source \#13, \#34 and \#38 are eps Eri (K2V star with G3.5~mag), $\pi^3$~Ori (F6V star with G3.2~mag), and $\beta$ Hy (G0V subgiant with G2.7~mag). We conclude that for FGK stars fainter than $\sim 3.5 \mbox{ mag}$, our broad--band flux values are essentially unaffected by optical loading.

\begin{figure}
   \centering
   \includegraphics[trim={0.5cm 0.0cm -0.1cm 0.0cm}, clip, width=0.48\textwidth,angle=0]{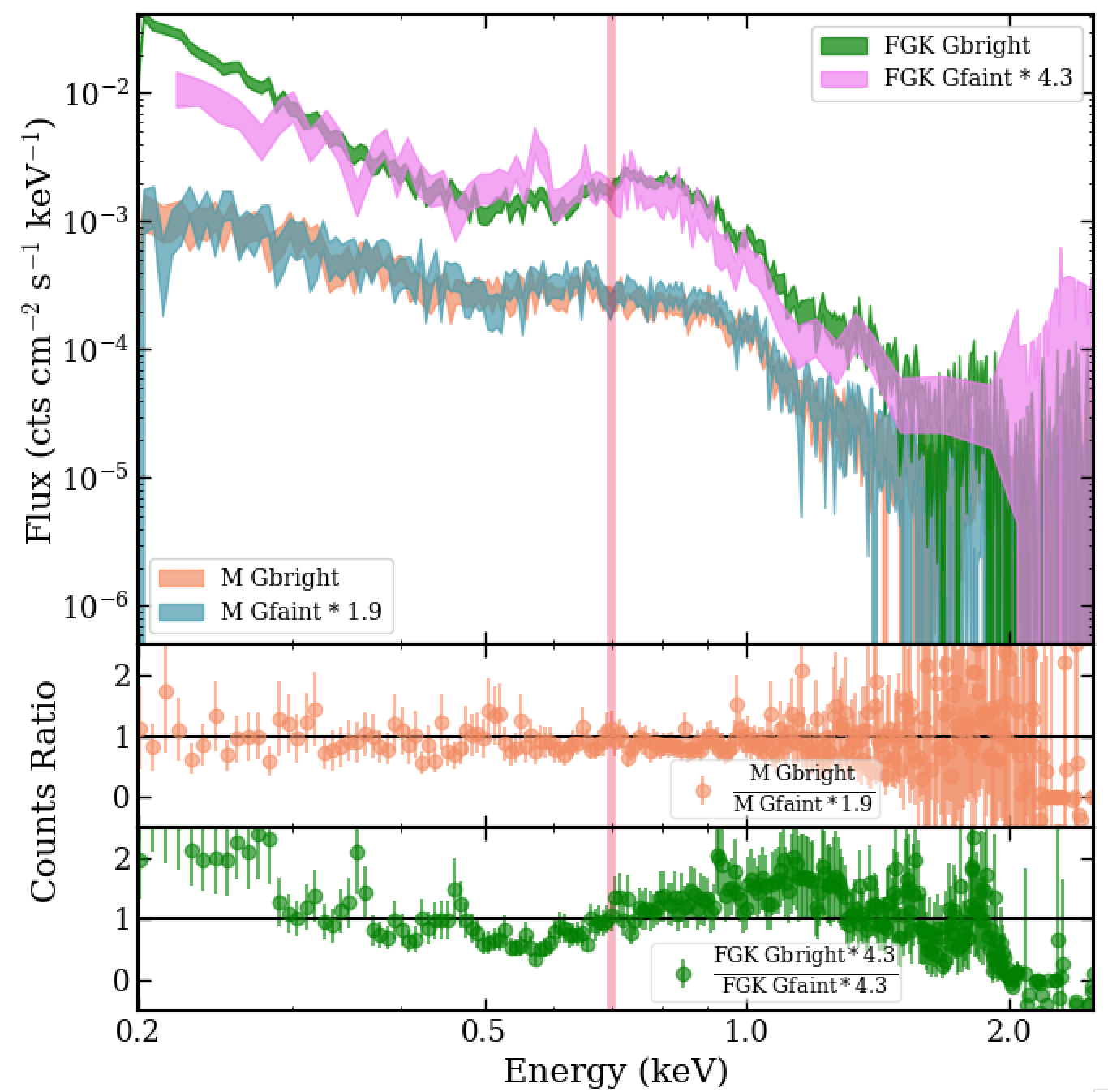}
   \caption{X-ray stacked spectra for the optically-bright M dwarf stars (orange: \( \mathrm{m}_G < 10\)), optically-faint (blue: \( \mathrm{m}_G > 10\)) M dwarf stars, optically-bright FGK dwarf stars (green: \( \mathrm{m}_G < 5\)) and optically-faint (violet: \( \mathrm{m}_G > 5\)) FGK dwarf stars. The middle and bottom panels display the ratio between the two groups of spectra.}
              \label{fig:optload_m}%
\end{figure}

\subsection{Optically bright and faint M stars}

In Section~\ref{sec:model_fitting}, we conclude the best-fit modeling of average M stars is 3T-APEC, though the softest component is with strangely low temperature (0.04 keV) and unconstrained normalization. 
Although optical loading was suspected initially for this soft component, all M dwarfs are fainter than the suggested optical loading threshold G = 5. 
The only two cases with companion stars brighter than G5 mag were removed from the sample.

Here we divide the M-dwarf sample into optically-bright and optically-faint groups using more conservative G-band magnitude cuts (G = 10). The counts spectra of two groups are presented in Figure \ref{fig:optload_m}, in the middle panels shows the ratio between optically-bright and -faint. 
All spectra are binned with minimal 20 counts in each channel. 
To facilitate the comparison, the spectrum of the M-Gfaint sample has been scaled up by a factor of 1.9 to match the flux level of the optically-bright M subset at 0.7 keV.
The ratio shows not soft excess for optically-bright groups and the two spectra are remarkable similar cross the band.
We therefore conclude that the M-dwarf sample is not affected by optical loading.
Instead, it likely originates from genuine emission that might be associated with deep coronal layers or cool regions.

\subsection{Optically bright and faint FGK stars}
To investigate the impact of optical loading, in Figure~\ref{fig:optload_m} we compare the stacked spectra of the 22 optically-bright FGK sample (green: G<5) with that of a subset of 9 less optically bright FGK stars (violet: G>5). 
For display it is scaled up by 4.3 times to match the optically-bright FGK subset at 0.7 keV (ratio normalized to 1 at 0.7 keV).

The ratio in the bottom panel of Figure~\ref{fig:optload_m} between the two spectra already reveals noticeable differences: below 0.35 keV, the G-bright spectrum shows a strong excess with a ratio exceeding 2.
We show as well the energy up to 2.5 keV for comparing. 
In addition, a slight mismatch around the emission lines above 1 keV may be related to the energy-shift effect of optical loading (as discussed in Effect 1 above).

We fit both spectra the 3T-APEC model in 0.2--2.5 keV. 
In Table~\ref{tab:fgk_ol2}, we list the best-fit parameters:
The fitted parameters for the optically bright and optically faint FGK stars are not consistent. The optically bright FGK group shows temperature and abundance values that are consistent with those of the full FGK sample, whereas the optically faint FGK group yields different temperature components.
Moreover, a very soft component is required to fit the optically bright FGK spectra but is not required in the optically faint group (unconstrained normalization).
This suggests that the optically bright FGK stars may dominate the average (all) FGK spectral shape and, ideally, should be excluded when aiming to isolate the impact of optical loading. However, the 10-pc optically faint FGK subset comprises only nine stars and therefore provides insufficient statistics. For such a small sample, fitting a three-temperature APEC model is likely to be driven by intrinsic source-to-source differences rather than by the effect of optical loading. This limitation highlights the need for a larger volume-limited sample in future work.

\begin{table}[h]
    \caption{Fitting parameters for optically bright and faint FGK stars.}\label{tab:fgk_ol2}
    \centering
    \begin{tabular}{l | ccc}
        \hline
        \toprule
        Param. &All FGK & Opt-bright FGK & Opt-faint FGK \\
         & & G<5 & G>5 \\
        \hline
        $kT_1$  & $0.072\pm0.003$ &$0.071 \pm 0.003$ & ${0.07}~{\rm fixed}$\\
        $N_1$   & ${ 33^{+11}_{-9}}$ &${56^{+16}_{-14}}$ & ${1^{+3}_{-1}}$\\
        $kT_2$  & ${ 0.36\pm0.02}$ & ${0.37}\pm 0.02$ & ${0.20}\pm 0.03$ \\
        $N_2$   &  $ 3.9\pm0.8$&${6} \pm 1 $ & ${3} \pm 1$ \\
        $kT_3$ & ${ 0.80\pm0.03}$ &${0.80} \pm 0.03$ & ${0.60} \pm 0.07$ \\
        $N_3$ &  ${ {2.4\pm0.6}}$&$5 \pm 1$ & $2 \pm 1$ \\
        Z& ${ 0.7^{+0.2}_{-0.1} }$&${0.6}\pm0.1$ & $0.4_{-0.2}^{ +0.4}$ \\
        \hline
        $\chi^2_\nu \frac{\chi^2}{d.o.f.}$ & ${ 1.4\frac{432.3}{301}}$ &${1.3\frac{272.32}{207}}$ & ${0.6\frac{79.76}{129}}$ \\
      \bottomrule
    \end{tabular}
    \tablefoot{The 3T-APEC model is used to fit in energy range 0.2--2.5 keV. Except for the all FGK spectrum, the channel is binned to have minimal 20 counts. $N_{\rm apec}$ is in the unit $10^{-4}\frac{{10}^{-14}}{4\pi(10 {\rm pc})^2}~ \mathrm{cm}^{-5}$.}
    
\end{table}

\subsection{Test of removing central optical `pile-up'}

To evaluate the impact of optical loading near the PSF core, we performed a test by excluding the central $5^{\prime\prime}$ region of all FGK and M dwarf sources prior to stacking their spectra. This method aims to reduce potential optical pile-up or loading effects concentrated in the image centre.

\begin{figure}[ht]
\centering
\includegraphics[trim={0.5cm 0.3cm 0.1cm 0.9cm}, clip, width=0.4\textwidth]{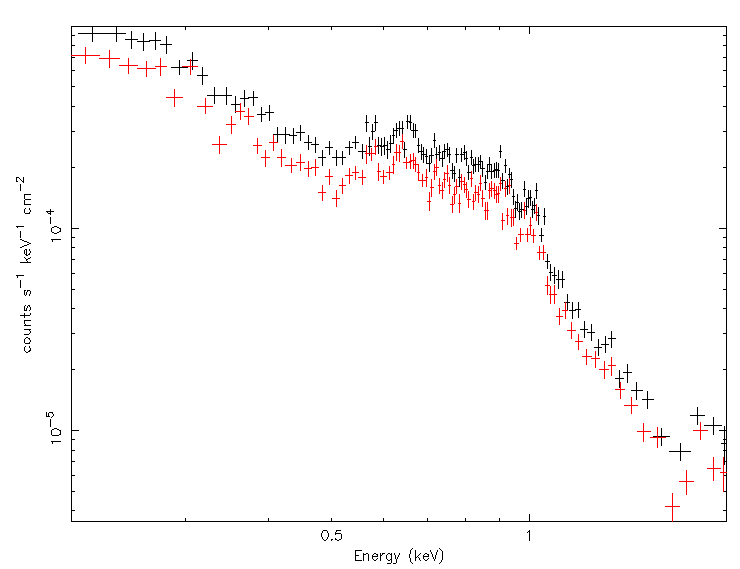}
\caption{Comparison between the original M stacked spectrum (black) and that obtained after excluding the central $5^{\prime\prime}$ region from each source (red).
\label{fig:MnoCen}}
\end{figure}

\begin{figure}[ht]
\centering
\includegraphics[trim={0.5cm 0.3cm 0.1cm 1.1cm}, clip, width=0.4\textwidth]{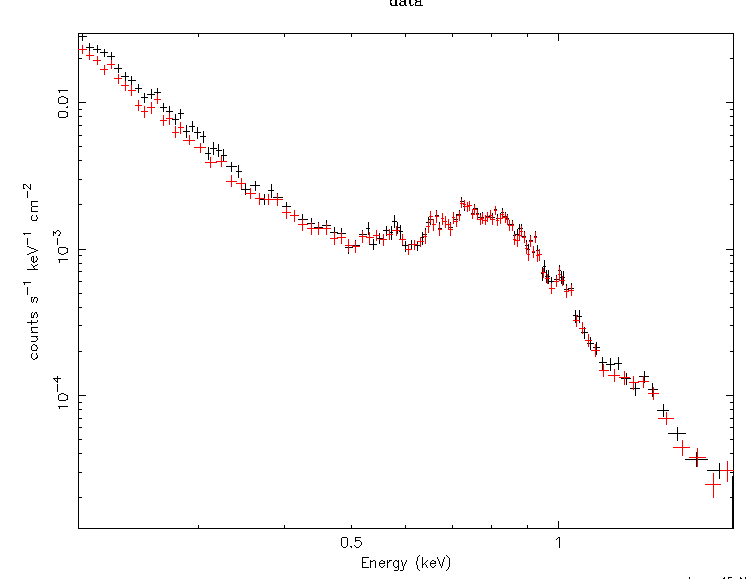}
\caption{Comparison between the original FGK stacked spectrum (black) and that obtained after excluding the central $5^{\prime\prime}$ region from each source (red).
\label{fig:FGKnoCen}}
\end{figure}

Figures~\ref{fig:MnoCen} and~\ref{fig:FGKnoCen} show a comparison between the original stacked spectra and those obtained after central-region removal, for FGK stars and M dwarfs, respectively.

For the FGK stars, we observe a $\sim20\%$ decrease in flux within the soft X-ray band (0.2--0.35~keV), while no significant (lower than 3\%) changes are seen at higher energies. Combined with the pattern-based tests indicating the presence of optical loading, this result suggests that the FGK sample is primarily affected in the soft band. In contrast, the M dwarf spectra is lowed by 25\% across the full energy range.

\section{Impact of different stacking method}\label{sec:app_stacking}
\begin{table}[h]
    \centering
    \caption{Best-fit parameters for M stars using different stacking strategies.}
    \begin{spacing}{1.2}
    \resizebox{\columnwidth}{!}{
    \begin{tabular}{l|cccc}
        \hline
        \toprule
	Param.	&	AC	&	AC T\_lt\_3000	&	AR 	&	AR T\_lt\_3000	\\
        \hline									

$kT_1$	&	$0.27 \pm 0.01$	&	$0.27 \pm 0.01$	&	$0.269 \pm 0.008$	&	$0.268 \pm 0.008$	\\
$N_1$	&	$1.7 \pm 0.2$	&	$1.8 \pm 0.2$	&	$1.5 \pm 0.1$	&	$1.5 \pm 0.1$	\\
$kT_2$	&	$0.94 \pm 0.03$	&	$0.95 \pm 0.03$	&	$0.97 \pm 0.03$	&	$0.97 \pm 0.02$	\\
$N_2$	&	$1.1 \pm 0.1$	&	$1.3^{+0.1}_{-0.2}$	&	$1.00 \pm 0.09$	&	$1.0 \pm 0.1$	\\
$Z$	&	$0.28\pm0.04$	&	$0.27\pm0.04$	&	$0.32\pm0.04$	&	$0.32\pm0.04$	\\
\addlinespace[2pt]
 \hline									
$\chi_{\nu}^2\ (\frac{\chi^2}{\mathrm{d.o.f.}})$	&	$0.93\frac{231.50}{249}$	&	$0.70\frac{172.63}{249}$	&	$2.2\frac{550.61}{249}$	&	$2.2\frac{559.26}{249}$	\\
\addlinespace[2pt]
\hline									
$L_\textrm{0.2-2.0}$	&	$2.6\pm0.1$	&	$2.89\pm0.06$	&	$2.53\pm0.04$	&	$2.58\pm0.04$	\\

\bottomrule
\end{tabular}}
\end{spacing}
\tablefoot{AC (Averaging Counts), AC\_T\_lt\_3000 (excluding sources with exposure $>3000$ s),  AR (Averaging Rates) and AC T\_lt\_3000 (excluding sources with exposure $>3000$ s). $N_{\rm apec}$ is in the unit $10^{-4}\frac{{10}^{-14}}{4\pi(10 {\rm pc})^2}~ \mathrm{cm}^{-5}$. Luminosities is in 0.2--2.0 keV with units of $10^{27}~\mathrm{erg\,s^{-1}}$ .}\label{tab:AC_AR}
\end{table}

\begin{table}[h]
    \centering
    \caption{Best-fit parameters for FGK stars using different stacking strategies.}
    \resizebox{\columnwidth}{!}{
    \begin{tabular}{l|cccc}
        \hline
        \toprule
Param.	&	AC	&	AC T\_lt\_1400	&	AR 	&	AR T\_lt\_1400	\\
 \hline									
$kT_1$	&	$0.09\pm0.01$	&	$0.101\pm0.009$	&	$0.097\pm0.007$	&	$0.102\pm0.008$	\\
$N_1$	&	$10^{+5}_{-3}$	&	$7^{+3}_{-2}$	&	$9^{+3}_{-2}$	&	$7\pm2$	\\
$kT_2$	&	$0.35\pm0.02$	& $0.36 \pm 0.02$	&	$0.35 \pm 0.01$	&	$0.35 \pm 0.01$	\\
$N_2$	&	$4.6 \pm 0.8$	&	$5.1\pm 0.9$	&	$5.5 \pm 0.9$	&	$6 \pm 1$	\\
$kT_3$	&	$0.79 \pm 0.03$	&	$0.79 \pm 0.03$	&	$0.81 \pm 0.03$	&	$0.81 \pm 0.02$	\\
$N_3$	&	$3.1 \pm 0.6$	&	$3.5 \pm 0.7$	&	$3.4 \pm 0.6$	&	$3.5 \pm 0.7$ \\
$Z$	&	$0.55^{+0.13}_{-0.09}$	&	$0.54^{+0.13}_{-0.09}$	&	$0.6\pm0.1$	&	$0.6\pm0.1$	\\
\addlinespace[2pt]
 \hline									
$\chi_{\nu}^2\ (\frac{\chi^2}{\mathrm{d.o.f.}})$	&	$1.15\frac{252.77}{219}$	&	$1.18\frac{257.44}{219}$	&	$2.10\frac{460.83}{219}$	&	$2.16\frac{472.60}{219}$	\\
\addlinespace[2pt] 
\hline									
$L_\textrm{0.35-2.0}$	& $12.1\pm0.2$ &	$13.0\pm0.2$	&	$14\pm1$	&	$15.3\pm0.5$	\\

\bottomrule
\end{tabular}}
\tablefoot{AC (Averaging Counts), AC\_T\_lt\_1400 (excluding sources with exposure $>3000$ s),  A (Averaging Rates) and AC T\_lt\_1400 (excluding sources with exposure $>1400$ s). Luminosities is in 0.35--2.0 keV with units of $10^{27}~\mathrm{erg\,s^{-1}}$ .}\label{tab:AC_AR_FGK}
\end{table}
Table~\ref{tab:AC_AR} and Table~\ref{tab:AC_AR_FGK} compare the spectral fitting results of the 3T-APEC model obtained using different stacking methods for M and FGK stars. Additionally, we also generated a version of the average spectrum excluding highly exposed sources (for M star is those with exposure times exceeding 3000 seconds, for FGK is 1400 seconds), referred to as T\_lt\_xxxx for both Averaging Counts (AC) method and Averaging Rates (AR) method.

The AR methods yield significantly larger reduced chi-squared values compared to those from the AC method. 
This is expected, as the error bars in the AR method are likely underestimated due to the assumption of Gaussian noise rather than Poisson statistics. 
Across all approaches, the best-fit temperatures and metallicities remain broadly consistent, indicating robustness in the overall spectral shape. 
The total luminosities obtained using the AC and AR methods are \(2.6 \times 10^{27}\) and \(2.53 \times 10^{27}\,\mathrm{erg\,s^{-1}}\) for M stars, and \(12.1 \times 10^{27}\) and \(14 \times 10^{27}\,\mathrm{erg\,s^{-1}}\) for FGK stars. 
The discrepancy is only seen for the FGK sample. We interpret it as a systematic effect arising from the choice of stacking procedure, likely amplified by intrinsic star-to-star differences within the sample, as discussed in Section~\ref{sec:method}.
When comparing luminosities, we find that excluding the highly exposed sources in the AC method does not significantly alter the spectral shape but increases the total luminosity by approximately 10\% in the 0.2--2.0\,keV band for M stars and 8\% for FGK stars.

\section{Spectrum extraction for individual sources} \label{sec:exctraction}
The extraction of source and background was carried out with \texttt{srctool} \citep{2022A&A...661A...1B}, with source and background radius adaptively chosen for detected source (\texttt{exttype}=AUTO) and fixed radius (source radius: 45", background annulus radius: 60", 99") for the undetected sources (\texttt{exttype=POINT}).
The fixed 45" sources radius is about 3 time of the \texttt{HEW} of SRG/\erosita\ survey mode. 
For the detected sources, the automatically determined source and background radii take into account the PSF, the detected counts, the background level, and the detection likelihood.
A few customized settings for the AUTO mode of \texttt{srctool} we used are listed in Table \ref{tab:srctool_autoreg_params} \footnote{See: \url{https://erosita.mpe.mpg.de/edr/DataAnalysis/srctool_doc.html}}.

\begin{table}
\caption{SRCTOOL autoregistration parameters.}
\label{tab:srctool_autoreg_params}
\centering
\resizebox{\columnwidth}{!}{
\begin{tabular}{ll}
\hline
\toprule
Parameter & Value \\
\hline
SRCTOOL\_AUTOREG\_MINIMUM\_SOURCE\_RADIUS & 45 \\
SRCTOOL\_AUTOREG\_INITIAL\_SRC\_R\_TO\_BACK\_R1 & 1.2 \\
SRCTOOL\_AUTOREG\_MAX\_SRC\_MAP\_TO\_BG\_MAP\_RATIO & 0.2 \\
SRCTOOL\_AUTOREG\_BACK\_TO\_SRC\_AREA\_RATIO & 3.0 \\
SRCTOOL\_AUTOREG\_MAX\_CONF\_MAP\_TO\_SRC\_MAP\_RATIO & 0.1 \\
SRCTOOL\_AUTOREG\_MAX\_RATIO\_BAC\_R1\_TO\_RADIUS\_99PC & 1 \\
SRCTOOL\_AUTOREG\_MAX\_BACK\_ANNULUS\_WIDTH & 120 \\
\bottomrule
\end{tabular}}
\end{table}

Amongst these settings, the most important is \nolinkurl{SRCTOOL_AUTOREG_BACK_TO_SRC_AREA_RATIO}, which defines the background region to be three times larger than the source region.
The detection status of the stars was determined by cross-matching with the \texttt{eRASS:1} catalogue \citep{Merloni2024AA}, selecting the nearest source within 15\arcsec, followed by visual inspection.
In cases where additional X-ray sources were present within the extraction region, they were masked using circular regions with radii equal to twice the \texttt{APERTURE} radius from the catalogue to minimize contamination from nearby sources.

\section{More configuration of elemental abundance for VAPEC}\label{sec:morevapec}
Here we show the result from fitting M dwarf stars with 1T-VAPEC and 2T-VAPEC (Table~\ref{tab:1vapec_moreA} and~\ref{tab:2vapec_moreA} respectively) using different initializations for the metal abundances. Parameters fixed to a number or to another parameter are shown by the $=$ symbol.
We adopt model C in the main text because it provides good fits for both the M- and FGK-star samples without imposing an arbitrary fixed He abundance or yielding unphysical parameter values. We note, however, that alternative model configurations may also be plausible.

\begin{table}
\centering
\caption{Fits of M dwarf with 1T-VAPEC using different initializations for the metal abundances.}
\begin{tabular}{llll}
\hline
\toprule
Param.	&	Model A	&	Model B	&	Model C 	\\
\hline
kT1     & $0.45^{+0.03}_{-0.02}$  & $0.46\pm0.02$ &   $0.46\pm0.02$ \\
N1      & $3.0\pm0.3$             & $3.6\pm0.2$&   $4.9\pm0.3$\\
A\_He   & =1& =1&   $0.10\pm0.04$ \\
A\_Rest & =A\_Ne                  & =A\_C&   =A\_He \\
A\_C    & =A\_Ne                  & $0.12\pm0.05$&   =A\_He \\
A\_O    & $0.23\pm0.05$           & $0.19\pm0.04$&   $0.12^{+0.04}_{-0.03}$ \\
A\_Ne   & $0.5\pm0.1$             & $0.63\pm0.06$&   $0.47\pm0.03$ \\
A\_Si   & $0.6\pm0.02$            & $0.38\pm0.2$&   $0.3\pm0.1$  \\
A\_Fe   & $0.09\pm0.02$           & $0.06\pm0.01$&  $0.041\pm0.008$ \\
\hline
$\chi^2$(d.o.f.)& 1.75 $\frac{433.75}{248}$   & 1.01 $\frac{249.48}{247}$& 1.03$\frac{255.33}{247}$ \\
\bottomrule
\end{tabular}
\tablefoot{The fitted energy range is 0.2--2.0 keV. 90\% uncertainty is reported. $N_{\rm apec}$ is in the unit $10^{-4}\frac{{10}^{-14}}{4\pi(10 {\rm pc})^2}~ \mathrm{cm}^{-5}$. The model C is the one used in Table~\ref{tab:Mfit}.}\label{tab:1vapec_moreA}
\end{table}

{\tiny
\begin{table}
\caption{Fits of FGK dwarf with 2T-VAPEC using different initializations for the metal abundances.}
\centering
\begin{spacing}{1.2}
\resizebox{\columnwidth}{!}{
\begin{tabular}{lllll}
\hline
\toprule
Param.	&	Model O	&	Model A  &  Model B	&	Model C 	\\
\hline
kT1     & $0.26^{+0.02}_{-0.03}$  & $0.26^{+0.2}_{-0.3}$ & $0.22^{+0.02}_{-0.01}$ & $0.22\pm0.01$\\
N1      & $10.2\pm4$              & $8\pm3$ &  $8\pm1$ & $10\pm2$ \\
KT2     & $0.74\pm0.06$           & $0.74\pm0.06$ &  $0.60\pm0.02$ & $0.60\pm0.02$\\
N2      & $6\pm1$                 & $5\pm0.9$ &  $7\pm 1$  & $8\pm2$ \\
A\_He   & =A\_Ne                   & =1 &  =1 & $0.3^{+0.2}_{-0.1}$\\
A\_Rest & =A\_Ne                   & =A\_Ne &  =A\_C &=A\_He\\
A\_C     & =A\_Ne                  & =A\_Ne &  $0.40^{+0.18}_{-0.12}$ &=A\_He\\
A\_O    & $0.18^{+0.14}_{-0.05}$   & $0.23^{+0.12}_{-0.06}$ &  $0.18^{+0.07}_{-0.05}$ & $0.14^{+0.06}_{-0.04}$\\
A\_Ne   & $0.4^{+0.4}_{-0.2}$ & $0.5^{+0.4}_{-0.2}$ &  $1.0\pm0.3$ & $0.8^{+0.3}_{-0.2}$\\
A\_Si   & $0.17^{+0.13}_{-0.09}$ & $0.2\pm0.1$ &  $0.2\pm0.1$ & $0.18^{+0.12}_{-0.09}$ \\
A\_Fe   & $0.35^{+0.11}_{-0.07}$ & $0.44^{+0.13}_{-0.07}$ & $0.35^{+0.09}_{-0.06}$ & $0.27^{+0.09}_{-0.06}$ \rule[-5pt]{0pt}{0pt}\\
\hline
$\chi^2$(d.o.f.)& 1.17 $\frac{264.25}{218}$   & 1.21 $\frac{264.53}{218}$ & 1.09 $\frac{235.81}{217}$ & 1.08$\frac{235.16}{217}$ \rule[-5pt]{0pt}{15pt} \\
\bottomrule
\end{tabular}
}
\tablefoot{The fitted energy range is 0.35--2.0 keV.  The model C here is the one used in Table~\ref{tab:FGKfit}.}\label{tab:2vapec_moreA}
\end{spacing}
\end{table}
}

\section{bootstrap resampling} \label{sec:bootstrap}

Besides the fit uncertainty, we perform a bootstrap resampling to estimate the uncertainty on the fit values by computing their scatter across the different realizations. 
This is a repeatable resampling from the full stars that form the same volume. 
In practice, we do 1000 realization. 
After stacking and fitting again every realization, we derive systematic uncertainties on each parameter by [5\%, 95\%] confidence intervals (90\% CI) from the obtained distribution of their values. 
The corresponding values are reported in the fit summary tables presented as Table~\ref{tab:Mfit} for M stars and Table~\ref{tab:FGKfit} for FGK stars.

\section{Luminosity in single eRASS}
\begin{figure}[]
    \centering
    \includegraphics[trim={0.1cm 0.1cm 0.1cm 0.1cm}, clip, width=\linewidth]{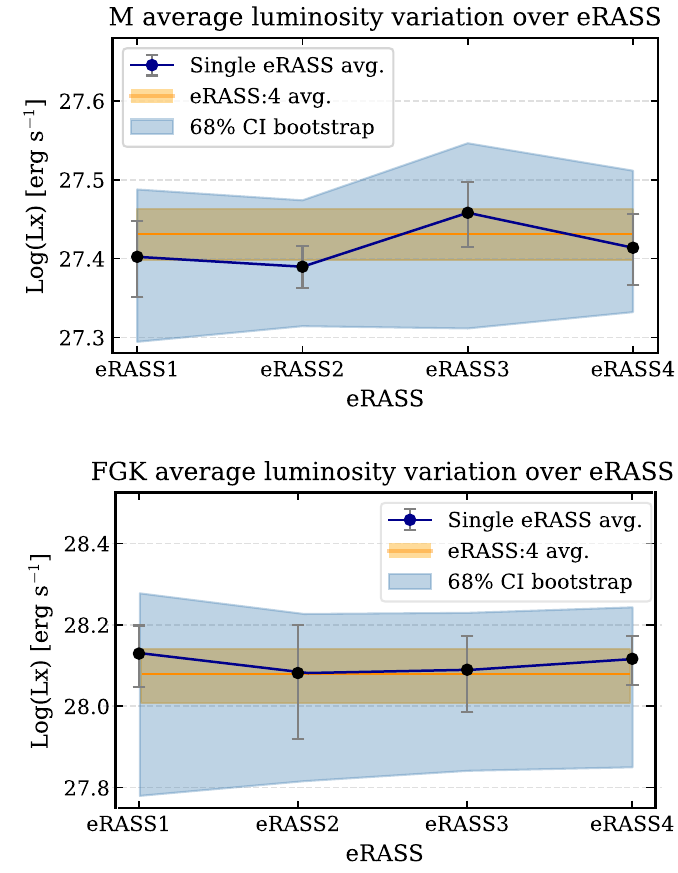}
    \caption{Average X-ray luminosity variation of nearby M (top) and FGK (bottom) dwarfs over the four eROSITA all-sky surveys (eRASS1 to eRASS4).
Blue points represent the average luminosity from individual eRASS epochs, with vertical bars indicating the methodological uncertainties. The orange horizontal band shows the average luminosity obtained from the combined eRASS:4 spectra, while the shaded blue area indicates the 68\% confidence interval estimated via bootstrap resampling.}
    \label{fig:separate_erass}
\end{figure}

Figure~\ref{fig:separate_erass} shows the variation in the average X-ray luminosity of nearby M (top) and FGK (bottom) dwarfs across the four eROSITA all-sky surveys (eRASS1 to eRASS4).
The overall consistency indicates that the mean luminosities derived for the 10-pc M- and FGK-dwarf samples defined in WGH are stable over time and not significantly affected by temporal variability.

\section{Assessing log-normal DEM models for M and FGK spectra}\label{sec:lognorm}

 We test to fit the average spectra with a multi-temperature plasma model that represents the emission as a log-normal distribution of APEC components, namely lognorm\footnote{See: \url{https://github.com/jeremysanders/lognorm}}. It approximates the differential emission measure by summing N discrete temperature components (default N=21) with temperatures evenly spaced in ln$T$. Each component is weighted by
$w_i \propto \exp\!\left[-\frac{1}{2}\left(\frac{\ln T_i-\ln T_{\rm c}}{\ln\sigma}\right)^2\right]$. As the width parameter decreases, the distribution approaches a single-temperature \texttt{apec} model. 
In Table~\ref{tab:lognorm}, we show that the single-\texttt{lognorm} fits improve upon the 1T-APEC and 1T-VAPEC models, but the overall fit quality remains poor. While the two-\texttt{lognorm} model provides acceptable fits for both stellar groups, it does not outperform the 2T-APEC model and yields poorly constrained width parameters (logsigma). We therefore find no statistically compelling evidence that \texttt{lognorm} models are required to describe the spectra.

A further insight from the single-\texttt{lognorm} fits is that they help explain the unrealistically low abundances obtained in the 1T case. For the M-star spectrum, the 1T-APEC fit yields Z=0.061, well below the $\sim$0.1--1$Z_\odot$ found in optical and infrared studies of local M dwarfs \citep{Montes2018MNRAS, Neves2012AA, Woolf2005MNRAS}.
This experiment with lognorm supports the interpretation that the extremely low metallicity is an artefact of modelling an intrinsically multi-temperature spectrum with an oversimplified 1T model.

{\tiny
\begin{table}
\caption{Fits using model \texttt{lognorm}.}
\centering
\begin{spacing}{1.2}
\resizebox{\columnwidth}{!}{
\begin{tabular}{lllll}
\hline
\toprule
Param.	&	M-1lognorm &	M2lognorm  &  FGK -1lognorm	& FGK-2lognorm	\\
\hline
kT1        & $0.41^{+0.02}_{-0.06}$   & $0.26\pm0.01$ & $0.44^{+0.02}_{-0.02}$ & $0.29\pm0.03$\\
logsigma   & $0.9^{+0.2}_{-0.1}$      & $0.08^{+0.07}_{-0.08}$ &  $0.55\pm0.06$ & $0.14^{+0.07}_{-0.14}$ \\
Z          & $0.24^{+0.03}_{-0.05}$   & $0.29^{+0.06}_{-0.05}$ &  $0.37^{+0.09}_{-0.06}$ & $0.34\pm0.06$\\
N1         & $3.1\pm0.3$              & $1.1\pm0.1$ &  $12\pm 1$  & $7\pm1$ \\
kT2        &                          & $0.94\pm0.03$ &   & $0.71^{+0.06}_{-0.05}$\\
N2         &                          & $1.6\pm0.2$ &   & $6.0^{+0.6}_{-1.0}$
\rule[-5pt]{0pt}{0pt}\\
\hline
$\chi^2$(d.o.f.)& 1.6 $\frac{405.45}{250}$   & 0.93 $\frac{230.70}{248}$  & 1.81 $\frac{403.13}{222}$ & 1.58$\frac{346.90}{220}$ \rule[0pt]{0pt}{5pt} \\
\bottomrule
\end{tabular}
}\label{tab:lognorm}
\end{spacing}
\end{table}
}

\section{Table of all stars}\label{sec:allstar}
Table~\ref{tab:M_alltable} and Table~\ref{tab:FGK_alltable} list the full set of stars included in our analysis and initially identified and catalogued by \citet{Caramazza2023AA}.
For each object, the table provides the identifier, coordinates (RA and DEC) at J2020, distance in parsecs, spectral type, \gaia\ G-band magnitude, and a flag indicating whether the source is detected in the \texttt{eRASS1 catalogue} (denoted as \texttt{IF\_ERASS}. 
The final column gives additional comments, including cross-matches with commonly known catalogue names. 
Distance in pc are computed as $1000 / \mathrm{PARALLAX}$ (mas), where the parallax values are taken from \gaia\ EDR3 \citep{Reyle2021AA}. Their 10 pc sample also includes stars slightly beyond 10 pc (e.g. 10.2 pc), whose parallax uncertainties make them statistically consistent with being within 10 pc.

\FloatBarrier
{\tiny
\onecolumn
\begin{longtable}{rlrrllrrrrr}
\caption{\gaia\ 10 pc M stars. Rate\_e1 denotes the eRASS1 catalogue count rate \citep{Merloni2024AA}, taken from the column \texttt{ML\_RATE\_0\_e1}, i.e., the 0.2--2.3 keV band rate. \textsuperscript{a}: Flag A - field crowded; Flag B - optical loading; Flag C- close pair. Detailed definition see Section~\ref{sec:sample}.}\label{tab:M_alltable}\\
\toprule
\textbf{ID} &                     
\textbf{OBJ\_NAME} &    
\textbf{RA\_mean} & 
\textbf{DEC\_mean} & 
\textbf{D (pc)}& 
\textbf{SpT} &    
\textbf{G (mag)}& 
\textbf{eRASS1 Catlog.} & 
\textbf{Gliese Name} &
\textbf{Comm.\textsuperscript{a}} & 
\textbf{Rate\_e1}\\

\midrule
\endfirsthead

\toprule
\textbf{ID} &                     
\textbf{OBJ\_NAME} &    
\textbf{RA\_mean} & 
\textbf{DEC\_mean} & 
\textbf{D (pc)}& 
\textbf{SpT} &    
\textbf{G (mag)}& 
\textbf{eRASS1 Catlog.} & 
\textbf{Gliese Name} &
\textbf{Comm.\textsuperscript{a}} & 
\textbf{Rate\_e1}\\
\midrule
\endhead
\midrule
\multicolumn{8}{r}{{Continued on next page}} \\
\midrule
\endfoot

\bottomrule
\endlastfoot
5	&	L 35-12	&	139.289739	&	-77.827856	&	$9.521\pm0.002$	&	M4.5V	&	11.5566	&	TRUE	&	GJ 1123	&	&	$0.05\pm0.01$\\
8	&	L 471-42	&	189.699638	&	-38.389215	&	$	6.663\pm0.001	$	&	M4.5V	&	11.2262	&	TRUE	&	GJ 3737	&	&$0.06\pm0.02$	\\
10	&	L 143-23	&	161.084287	&	-61.200422	&	$	4.832\pm0.001	$	&	M5.5V	&	11.8594	&	FALSE	&	GJ 3618	& &	-	\\
17	&	AD Leonis	&	154.898084	&	19.869750	&	$	4.965\pm0.001	$	&	M3V	&	8.2040	&	TRUE	&	GJ 388	&	& $25.50\pm0.60$	\\
18	&	GJ 832	&	323.391156	&	-49.013754	&	$	4.967\pm0.001	$	&	M2V	&	7.7399	&	TRUE	&	GJ 832	&	& $0.14\pm0.05$ \\
20	&	L 49-19	&	343.915846	&	-75.464834	&	$	8.597\pm0.001	$	&	M3V	&	9.2927	&	TRUE	&	GJ 877	&	& $0.13\pm0.03$	\\
21	&	UCAC4 195-119117	&	235.199442	&	-51.028548	&	$	5.327\pm0.001	$	&	M6V	&	12.7359	&	TRUE	&	&& $0.07\pm0.02$	\\
23	&	GJ 176	&	70.736437	&	18.951667	&	$	9.485\pm0.002	$	&	M2V	&	9.0044	&	TRUE	&	GJ 176	&	& $0.06\pm0.02$	\\
28	&	HD 191849	&	303.478909	&	-45.164951	&	$	6.165\pm0.001$  &	M0.5V&	7.2426	&	TRUE	&	GJ 784	&   & $0.25\pm0.06$	\\
32	&	GJ 581	&	229.854606	&	-7.722841	&	$	6.300\pm0.001	$	&	M3V	&	9.4217	&	FALSE	&	GJ 581	&	& -	\\
33	&	CD-51 6859	&	189.457785	&	-52.001305	&	$	9.492\pm0.002	$	&	M3V	&	9.5650	&	TRUE	&	GJ 479	& & $0.88\pm0.07$\\
35	&	GJ 570 B	&	224.366845	&	-21.421181	&	$	5.9\pm0.8	$	&	M1.5V	&	7.2492	&	TRUE	&	GJ 570 A&Flag A& $1.19\pm0.09$	\\
42	&	G 42-24	&	148.477730	&	20.948620	&	$	9.545\pm0.003	$	&	M4.5V	&	12.3635	&	TRUE	&	GJ 3571	& &	$0.20\pm0.05$	\\
45	&	HD 50281 Ba	&	103.071805	&	-5.190088	&	$	8.750\pm0.002	$	&	M2.5V	&	9.0975	&	TRUE	&	Flag A 	&&$0.21\pm0.06$	\\
47	&	CD-45 5378	&	146.120459	&	-45.779899	&	$	9.419\pm0.001	$	&	M2.5V	&	9.1587	&	TRUE	&	GJ 367	&	&$0.06\pm0.02$	\\
49	&	WT 460 A	&	212.994105	&	-41.539815	&	$	9.11\pm0.02	$	&	M5.5V	&	13.4190	&	TRUE	&		&	&$0.21\pm0.04$	\\
51	&	GJ 667 C	&	259.753153	&	-34.998092	&	$	7.243\pm0.001	$	&	M2V	&	9.3861	&	FALSE	&		&	Flag A & -	\\
65	&	Wolf 358	&	162.711778	&	6.803363	&	$	6.967\pm0.001	$	&	M4V	&	10.2907	&	FALSE	&	GJ 402	&	&-	\\
66	&	BD-18 359 A	&	31.278116	&	-17.615637	&	$	9.32\pm0.02	$	&	M2.5V	&	9.1705	&	TRUE	&	GJ 84 A	&	&$0.10\pm0.03$	\\
68	&	GJ 486	&	191.979981	&	9.748721	&	$	8.079\pm0.002	$	&	M3.5V	&	10.1051	&	FALSE	&	GJ 486	&	&-	\\
69	&	CD-68 47 A	&	17.601313	&	-67.441587	&	$	7.88\pm0.02	$	&	M3V	&	8.7420	&	TRUE	&	GJ 54 A	&	&$0.12\pm0.03$	\\
71	&	L 674-15	&	123.170455	&	-21.555979	&	$	8.117\pm0.002	$	&	M4V	&	10.6716	&	FALSE	&	GJ 300	&	&	-\\
73	&	G 13-22	&	183.563417	&	0.622336	&	$	8.088\pm0.003	$	&	M5V	&	11.8286	&	TRUE	&	GJ 1154	&	& $0.55\pm0.07$	\\
74	&	HD 95735	&	165.829974	&	35.942082	&	$	2.546\pm0.000	$	&	M2V	&	6.5511	&	TRUE	&	GJ 411	&	&	$1.04\pm0.11$\\
75	&	Kapteyn's Star	&	77.972578	&	-45.051665	&	$	3.934\pm0.000	$	&	M0.5V	&	8.0635	&	TRUE	&	GJ 191	&	& $0.04\pm0.01$	\\
76	&	Ross 64	&	96.175515	&	23.430079	&	$	8.494\pm0.002	$	&	M4V	&	11.6625	&	FALSE	&	GJ 232	&	&- \\
77	&	GJ 1061	&	54.004832	&	-44.514876	&	$	3.674\pm0.000	$	&	M5.5V	&	10.9954	&	TRUE	&	GJ 1061	&	& $0.05\pm0.01$ 	\\
78	&	BD+01 2447	&	157.227954	&	0.836740	&	$	7.038\pm0.001	$	&	M2V	&	8.6759	&	TRUE	&	GJ 393	&	& $0.15\pm0.05$\\
79	&	Ross 837	&	209.556042	&	12.583054	&	$	9.360\pm0.004	$	&	M3.5V	&	11.0511	&	TRUE	&	GJ 3817	&	& $0.04\pm0.02$\\
80	&	AN Sex	&	153.072727	&	-3.747083	&	$	7.707\pm0.001	$	&	M2V	&	8.3317	&	TRUE	&	GJ 382	&	&	$0.54\pm0.08$\\
81	&	Proxima Cen	&	217.380988	&	-62.675016	&	$	1.302\pm0.000	$	&	M5.5V	&	8.9847	&	TRUE	&	GJ 551	&	&	$3.61\pm0.16$\\
86	&	G 109-35	&	104.875677	&	19.343643	&	$	7.753\pm0.004	$	&	M5.5V	&	12.7764	& FALSE	&	GJ 1093	&	&	-\\
89	&	Wolf 424 B	&	188.312294	&	9.022109	&	$	4.475\pm0.009	$	&	M5V	&	11.2401	&	TRUE	&	GJ 473 B	&	FLAG C&	$2.21\pm0.15$\\
91	&	SCR J1138-7721	&	174.514990	&	-77.359910	&	$	8.379\pm0.002	$	&	M5.5V	&	12.8162	&	FALSE	&		&	&	-\\
95	&	PM J11413-3624	&	175.343179	&	-36.407786	&	$	8.689\pm0.003	$	&	M4V	&	11.5837	&	FALSE	&		&	&	-\\
96	&	LP 991-84	&	24.841623	&	-39.603823	&	$	8.726\pm0.003	$	&	M5V	&	12.5683	&	FALSE	&		&	&	-\\
97	&	Ross 47	&	85.550522	&	12.480201	&	$	5.791\pm0.001	$	&	M4V	&	10.1137	&	FALSE	&	GJ 213	&	&	-\\
99	&	CD-40 5404	&	144.939134	&	-41.065479	&	$	9.607\pm0.002	$	&	M3V	&	9.6314	&	TRUE	&	GJ 358	&	&	$1.34\pm0.12$\\
107	&	SCR J0740-4257	&	115.045352	&	-42.958254	&	$	7.981\pm0.001	$	&	M5V	&	12.0307	&	TRUE	&		&	&	$0.67\pm0.07$\\
108	&	Ross 695	&	186.225477	&	-18.255733	&	$	8.875\pm0.002	$	&	M2.5V	&	10.2900	&	FALSE	&	GJ 465	&	&	-\\
110	&	G 100-28 A	&	85.107758	&	24.800031	&	$	10.2\pm0.3	$	&	M5.5V	&	13.3290	&	TRUE	&	GJ 1083 A	&	&	$0.37\pm0.07$\\
117	&	LP 776-46	&	75.832293	&	-17.376122	&	$	9.236\pm0.003	$	&	M3V	&	10.5616	&	FALSE	&	GJ 3325	&	& -	\\
119	&	G 112-50	&	117.979331	&	-0.007810	&	$	9.272\pm0.004	$	&	M4.5V	&	11.6534	&	FALSE	&	GJ 1103 A	&	&	-\\
120	&	HD 32450 A	&	75.617585	&	-21.258249	&	$	8.363\pm0.003	$	&	M0.5V	&	7.7924	&	TRUE	&	GJ 185 A	&	FLAG C& $0.12\pm0.03$\\
122	&	G 41-14 B	&	134.736998	&	8.472024	&	$	6.77\pm0.09	$	&	M4V	&	10.3290	&	TRUE	&	GJ 3522 B	&	FLAG C  &	$6.71\pm0.29$\\
124	&	BD-17 588 A	&	45.461893	&	-16.594901	&	$	6.864\pm0.001	$	&	M3V	&	10.0584	&	TRUE	&	GJ 3193	&	FLAG C	&  $1.67\pm0.10$\\
125	&	GJ 357	&	144.007690	&	-21.666563	&	$	9.436\pm0.002	$	&	M2.5V	&	9.8916	&	FALSE	&	GJ 357	&	&	-\\
130	&	GJ 251	&	103.698934	&	33.265860	&	$	5.585\pm0.001	$	&	M3V	&	8.8635	&	TRUE	&	GJ 251	&	&	$0.10\pm0.04$\\
131	&	L 347-14	&	290.205411	&	-45.575090	&	$	5.909\pm0.002	$	&	M4V	&	10.7529	&	FALSE	&	GJ 754	&	&	-\\
132	&	41 Ara B	&	259.770440	&	-46.636169	&	$	8.83\pm0.03	$	&	M0.5V	&	7.9870	&	FALSE	&	GJ 666 B	&	&	-\\
133	&	Ross 619	&	122.996172	&	8.743393	&	$	6.769\pm0.004	$	&	M4V	&	11.3972	&	FALSE	&	GJ 299	&	&-	\\
134	&	L 173-19	&	30.160714	&	-55.968356	&	$	8.216\pm0.002	$	&	M3.5V	&	10.5571	&	TRUE	&		&	&	$0.06\pm0.02$ \\
135	&	G 113-20	&	124.031065	&	1.302926	&	$	8.939\pm0.002	$	&	M2V	&	9.1331	&	FALSE	&	GJ 2066	&	& -	\\
136	&	BD-17 588 B	&	45.461893	&	-16.594901	&	$	6.864\pm0.001	$	&	M3.5V	&	10.4310	&	TRUE	&	GJ 3193 B	&	FLAG C	& $1.67\pm0.10$ \\
137	&	HD 102365 B	&	176.624462	&	-40.494333	&	$	8.028\pm0.003	$	&	M4V	&	11.7247	&	TRUE	&	GJ 442 B	&	FLAG B&	$0.03\pm0.01$\\
138	&	CD-44 3045 A	&	104.435181	&	-44.291463	&	$	9.309\pm0.003	$	&	M3V	&	10.4182	&	TRUE	&	GJ 257 A	&	FLAG C	& $0.04\pm0.02$\\
139	&	CD-40 9712	&	233.044774	&	-41.281592	&	$	5.917\pm0.001	$	&	M2.5V	&	8.2741	&	TRUE	&	GJ 588	&	&	$0.28\pm0.06$\\
140	&	G 41-14 A	&	134.736866	&	8.472006	&	$	6.77\pm0.09	$	&	M4V	&	9.97131	&	TRUE	&	GJ 3522 A	&	FLAG C	& $6.71\pm0.29$\\
141	&	HD 260655	&	99.290328	&	17.566781	&	$	9.998\pm0.002	$	&	M0.5V	&	8.8782	&	TRUE	&	GJ 239	&	&	$0.07\pm0.03$\\
142	&	GJ 273	&	111.855418	&	5.204304	&	$	3.786\pm0.001	$	&	M3.5V	&	8.5763	&	TRUE	&	GJ 273	&	&	$0.11\pm0.04$\\
143	&	Ross 128	&	176.938524	&	0.797437	&	$	3.375\pm0.000	$	&	M4V	&	9.6010	&	TRUE	&	GJ 447	&	&	$0.20\pm0.05$\\
144	&	G 9-38 A	&	134.558068	&	19.762822	&	$	5.151\pm0.003	$	&	M5.5V	&	11.9662	&	TRUE	&	GJ 1116 A	&	FLAG C &$0.94\pm0.12$	\\
145	&	G 161-7 A	&	138.899291	&	-10.597554	&	$	9.68\pm0.09	$	&	M5V	&	12.0050	&	TRUE	&		&	&	$0.46\pm0.07$ \\
146	&	Wolf 424 A	&	188.311992	&	9.022391	&	$	4.327\pm0.010	$	&	M5.5V	&	11.2351	&	TRUE	&	GJ 473 A	&	FLAG C&	$2.21\pm0.15$\\
147	&	Ross 104	&	165.015041	&	22.831314	&	$	6.748\pm0.001	$	&	M2.5V	&	8.9770	&	TRUE	&	GJ 408	&	&	$0.15\pm0.05$\\
148	&	L 403-31	&	210.961771	&	-42.699747	&	$	9.940\pm0.003	$	&	M4V	&	11.6951	&	TRUE	&		&	&	$0.98\pm0.08$\\
149	&	G 99-49	&	90.016402	&	2.706319	&	$	5.208\pm0.001	$	&	M4V	&	9.9013	&	TRUE	&	GJ 3379	&	&	$2.99\pm0.18$\\
150	&	L 100-115	&	145.691871	&	-68.878432	&	$	6.504\pm0.001	$	&	M4.5V	&	11.1332	&	TRUE	&	GJ 1128	&	&	$0.08\pm0.01$\\
151	&	L 32-9	&	98.423335	&	-75.628355	&	$	8.839\pm0.001	$	&	M3V	&	9.3847	&	TRUE	&		&	FLAG C	& $0.19\pm0.02$\\
152	&	L 32-8	&	98.437148	&	-75.623442	&	$	8.839\pm0.001	$	&	M3.5V	&	10.2122	&	TRUE	&		&	FLAG C	& $0.19\pm0.02$\\
153	&	Ross 614 A	&	97.351828	&	-2.818235	&	$	4.12\pm0.01	$	&	M4.5V	&	9.6300	&	TRUE	&	GJ 234 A	& &	 $5.66\pm0.34$\\
154	&	CD-48 11837 B	&	263.805929	&	-48.677114	&	$	9.690\pm0.004	$	&	M3.5V	&	11.6138	&	TRUE	&		&	FLAG C & $0.12\pm0.04$	\\
155	&	BD+11 2576	&	202.505778	&	10.370912	&	$	7.628\pm0.002	$	&	M1V	&	8.2052	&	FALSE	&	GJ 514	&	&-\\
156	&	SCR J0630-7643 A	&	97.692710	&	-76.716283	&	$	8.876\pm0.007	$	&	M5.5V	&	13.2158	&	TRUE	&		&	& $0.29\pm0.02$\\
157	&	GJ 166 C	&	63.826537	&	-7.675594	&	$	5.014\pm0.002	$	&	M4V	&	9.77531	&	TRUE	&	GJ 166 C	&	FLAG B	& $5.47\pm0.22$\\
158	&	GJ 229 A	&	92.643378	&	-21.868812	&	$	5.761\pm0.001	$	&	M1V	&	7.3134	&	TRUE	&	GJ 229 A	&	& $0.36\pm0.06$\\
159	&	HD 225213	&	1.393036	&	-37.370956	&	$	4.346\pm0.001	$	&	M2V	&	7.6824	&	TRUE	&	GJ 1	&	&	$0.07\pm0.03$ \\
160	&	CD-37 10765 B	&	245.008169	&	-37.523965	&	$	8.498\pm0.008	$	&	M5V	&	12.8980	&	TRUE	&	GJ 618 b	&	FLAG C	&  $0.06\pm0.02$\\
161	&	AP Col	&	91.217468	&	-34.557941	&	$	8.666\pm0.002	$	&	M5V	&	11.1121	&	TRUE	&		&	&	$1.93\pm0.12$\\
162	&	GJ 3323	&	75.486041	&	-6.949322	&	$	5.375\pm0.001	$	&	M4V	&	10.6493	&	TRUE	&	GJ 3323	&	&	$0.65\pm0.07$\\
163	&	Wolf 359	&	164.097831	&	6.999014	&	$	2.409\pm0.000	$	&	M6V	&	11.0383	&	TRUE	&	GJ 406	&&	$1.65\pm0.16$	\\
164	&	GL Vir	&	184.739967	&	11.127232	&	$	6.464\pm0.002	$	&	M5.5V	&	11.9261	&	TRUE	&	GJ 1156	&&	$0.60\pm0.08$	\\
165	&	G 9-38 B	&	134.557322	&	19.762531	&	$	5.095\pm0.005	$	&	M6V	&	12.4856	&	TRUE	&	GJ 1116 B	&	FLAG C & $0.94\pm0.12$	\\
166	&	L 737-9 A	&	77.149104	&	-18.180199	&	$	9.231\pm0.008	$	&	M3.5V	&	9.0548	&	TRUE	&	GJ 190 A	&	&$0.07\pm0.02$	\\
167	&	G 89-32 A	&	114.106052	&	7.076866	&	$	8.58\pm0.07	$	&	M5V	&	12.0130	&	TRUE	&	GJ 3454	&&	$0.92\pm0.12$	\\
168	&	CD-48 11837 A	&	263.807391	&	-48.678131	&	$	9.680\pm0.002	$	&	M2V	&	9.29361	&	TRUE	&	GJ 680	&	FLAG C & $0.12\pm0.04$	\\
169	&	CD-37 10765 A	&	245.009125	&	-37.523251	&	$	8.513\pm0.002	$	&	M3V	&	9.49538	&	TRUE	&	GJ 618 A	&	FLAG C&	$0.06\pm0.02$\\
170	&	GJ 674	&	262.171317	&	-46.900316	&	$	4.553\pm0.001	$	&	M3V	&	8.3364	&	TRUE	&	GJ 674	&	&	$1.72\pm0.13$\\
171	&	GJ 433	&	173.861792	&	-32.544918	&	$	9.077\pm0.002	$	&	M2V	&	8.8967	&	TRUE	&	GJ 433	&	&  $0.05\pm0.02$	\\
177	&	HD 36395	&	82.868591	&	-3.689406	&	$	5.704\pm0.001	$	&	M1.5V	&	7.1082	&	TRUE	&	GJ 205	&	&	$0.59\pm0.08$\\
183	&	L 768-119 A	&	235.514844	&	-19.477795	&	$	9.69\pm0.03	$	&	M3V	&	10.6858	&	FALSE	&	GJ 595 A	&&	-	\\
184	&	G 48-20	&	142.682461	&	0.319467	&	$	9.910\pm0.003	$	&	M3.5V	&	10.5017	&	FALSE	&	GJ 1125	&&		-\\
185	&	UPM J0815-2344 A	&	123.797365	&	-23.737305	&	$	9.897\pm0.003	$	&	M3.5V	&	11.0593	&	TRUE	&		&	&	$0.08\pm0.04$\\
186	&	GJ 682	&	264.259530	&	-44.324673	&	$	5.008\pm0.001	$	&	M4V	&	9.6021	&	TRUE	&	GJ 682	&	&	$0.10\pm0.04$\\
187	&	G 119-36 A	&	162.435102	&	35.541427	&	$	9.73\pm0.04	$	&	M4V	&	11.6252	&	FALSE	&	GJ 1138 A	&	&	-\\
188	&	GJ 436	&	175.552054	&	26.701835	&	$	9.775\pm0.003	$	&	M3V	&	9.5820	&	FALSE	&	GJ 436	&	&	-\\
191	&	HD 119850	&	206.443094	&	14.883051	&	$	5.435\pm 0.001	$	&	M1.5V	&	7.6105	&	TRUE	&	GJ 526	&	&	$0.13\pm0.03$\\
195	&	YZ CMi	&	116.165357	&	3.549871	&	$	5.989\pm0.001	$	&	M4V	&	9.6922	&	TRUE	&	GJ 285	&	&	$9.67\pm0.36$\\
196	&	CD-44 3045 B	&	104.434219	&	-44.291824	&	$	8.041\pm0.003	$	&	M3V	&	10.3819	&	TRUE	&	GJ 257 B	&	FLAG C&	$0.04\pm0.02$\\
197	&	BD-11 3759	&	218.567931	&	-12.516108	&	$	6.253\pm0.002	$	&	M4V	&	9.8949	&	FALSE	&	GJ 555	&	&	-\\
199	&	G 160-28	&	57.681931	&	-6.102877	&	$	9.928\pm0.004	$	&	M3.5V	&	11.4772	&	FALSE	&	GJ 1065	&	&	-\\
205	&	Ross 440 A	&	142.835243	&	-13.488438	&	$	10.0\pm0.4	$	&	M3V	&	9.6544	&	FALSE	&	GJ 352 A	&	FLAG C&	-\\
206	&	Ross 440 B	&	142.835386	&	-13.488438	&	$	10.0\pm0.4	$	&	M2.5V	&	9.8487	&	FALSE	&	GJ 352 B	&	FLAG C &	-\\
207	&	L 399-68	&	190.186590	&	-43.562373	&	$	8.052\pm0.002	$	&	M3.5V	&	10.9834	&	TRUE	&	GJ 9415	&&		$0.08\pm0.02$\\
208	&	L 230-188	&	62.609087	&	-53.616320	&	$	7.108\pm0.001	$	&	M4.5V	&	11.9028	&	TRUE	&	GJ 1068	&	Flag A&	$0.03\pm0.01$\\
210	&	HD 32450 B	&	75.617601	&	-21.257964	&	$	8.416\pm0.006	$	&	M0V	&	9.63678	&	TRUE	&	GJ 185 B	&	FLAG C&	$0.12\pm0.03$ \\
211	&	Wolf 461	&	195.134174	&	5.686880	&	$	8.545\pm0.004	$	&	M4.5V	&	11.7729	&	TRUE	&	GJ 493.1	&&	$0.68\pm0.08$	\\
217	&	L 205-128	&	266.630598	&	-57.326924	&	$	5.889\pm0.002	$	&	M3V	&	9.5922	&	FALSE	&	GJ 693	&&	-	\\
223	&	CD-30 731	&	31.448775	&	-30.176035	&	$	9.380\pm0.002	$	&	M3V	&	11.0506	&	TRUE	&	GJ 3135	&	&	$0.32\pm0.05$ \\

\end{longtable}

\begin{longtable}{rlrrllrrrrr}
\caption{\gaia\ 10 pc FGK stars. Rate\_e1 denotes the eRASS1 catalogue count rate \citep{Merloni2024AA}, taken from the column \texttt{ML\_RATE\_0\_e1}, i.e., the 0.2--2.3 keV band rate.  Flag A: field crowded; Flag B: optical loading; Flag C: close pair.}\label{tab:FGK_alltable} \\
\toprule
\textbf{ID} &                     
\textbf{OBJ\_NAME} &    
\textbf{RA\_mean} & 
\textbf{DEC\_mean} & 
\textbf{D (pc)}& 
\textbf{SpT} &    
\textbf{G (mag)}& 
\textbf{eRASS1 Catlog.} & 
\textbf{Gliese Name} &
\textbf{Comm.} & 
\textbf{Rate\_e1}\\

\midrule
\endfirsthead

\toprule
\textbf{ID} &                     
\textbf{OBJ\_NAME} &    
\textbf{RA\_mean} & 
\textbf{DEC\_mean} & 
\textbf{D (pc)}& 
\textbf{SpT} &    
\textbf{G (mag)}& 
\textbf{eRASS1 Catlog.} & 
\textbf{Gliese Name} &
\textbf{Comm.} &
\textbf{Rate\_e1}\\
\midrule
\endhead
\midrule
\multicolumn{8}{r}{{Continued on next page}} \\
\midrule
\endfoot
\bottomrule
\endlastfoot
0	&	GJ 570 A	&	224.3731	&	-21.4255	&	$5.886\pm0.002$	&	K4V	&	5.3640	&	TRUE	&	GJ 570 A	&	Flag A	&	$0.98\pm0.05$\\
2	&	41 Ara Aa	&	259.774717	&	-46.635611	&	$8.791\pm0.006$	&	G9V	&	5.3014	&	FALSE	&	GJ 666 A	&	&	$0.05\pm0.01$	\\
6	&	del Eri	&	55.811533	&	-9.759060	&	$9.09\pm0.02$	&	K0V	&	3.2755	&	TRUE	&	GJ 150	&	Flag B	&	$0.60\pm0.03$\\
9	&	61 UMa	&	175.262480	&	34.199415	&	$9.576\pm0.009$	&	G8V	&	5.1109	&	TRUE	&	GJ 434	&	&	$2.03\pm0.08$	\\
10	&	61 Vir	&	199.594747	&	-18.317386	&	$8.53\pm0.01$	&	G6.5V	&	4.5325	&	TRUE	&	GJ 506	&	Flag B	&	$0.09\pm0.01$\\
12	&	gam Pav	&	321.611982	&	-65.361539	&	$9.258\pm0.009$	&	F9V	&	4.0891	&	TRUE	&	GJ 827	&	Flag B	&	$0.05\pm0.01$\\
13	&	eps Eri	&	53.226934	&	-9.458139	&	$3.220\pm0.001$	&	K2V	&	3.4657	&	TRUE	&	GJ 144	&	Flag B	& $14.63\pm0.20$	\\
15	&	del Pav	&	302.199169	&	-66.188646	&	$6.099\pm0.005$	&	G8V	&	3.3641	&	TRUE	&	GJ 780	&	Flag B	&	$0.41\pm0.03$\\
17	&	HD 20794	&	50.006059	&	-43.065552	&	$6.041\pm0.003$	&	G6V	&	4.0639	&	TRUE	&	GJ 139	&	Flag B	&	$0.09\pm0.01$\\
25	&	36 Oph B	&	258.834400	&	-26.609468	&	$5.952\pm0.005$	&	K1V	&	4.8356	&	TRUE	&	GJ 663 B	&	Flag B	C & $3.31\pm0.09$	\\
26	&	GJ 216 B	&	86.108650	&	-22.423890	&	$8.892\pm0.002$	&	K2.5V	&	5.8681	&	TRUE	&	GJ 216 B	&	&	$1.01\pm0.04$	\\
28	&	HD 50281 A	&	103.072033	&	-5.173733	&	$8.745\pm0.003$	&	K3.5V	&	6.2310	&	TRUE	&	GJ 250 A	&	&	$0.74\pm0.05$	\\
32	&	36 Oph A	&	258.834400	&	-26.609468	&	$5.948\pm0.004$	&	K2V	&	4.8285	&	TRUE	&	GJ 663 A	&	Flag B	 C	& $3.31\pm0.09$\\
34	&	pi.03 Ori	&	72.462767	&	6.961347	&	$8.02\pm0.01$	&	F6V	&	3.0879	&	TRUE	&	GJ 178	&	Flag B	&	$12.63\pm0.17$\\
35	&	eps Ind A	&	330.882374	&	-56.800743	&	$3.638\pm0.001$	&	K5	&	4.3229	&	TRUE	&	GJ 845 A	&	Flag B	&	$1.36\pm0.06$\\
36	&	GJ 66 A	&	24.951421	&	-56.196385	&	$8.196\pm0.002$	&	K2V	&	5.6254	&	TRUE	&	GJ 66 A	&	GJ 66 B	 C&	$0.52\pm0.03$\\
38	&	bet Hyi	&	6.496340	&	-77.252346	&	$7.48\pm0.02$	&	G0V	&	2.6807	&	TRUE	&	GJ 19	&	Flag B	&	$2.95\pm0.07$\\
39	&	HD 102365 A	&	176.617751	&	-40.498010	&	$9.319\pm0.008$	&	G2V	&	4.7016	&	TRUE	&	GJ 442 A	&	Flag B&	 $0.04\pm0.01$	\\
41	&	GJ 216 A	&	86.113957	&	-22.450530	&	$8.91\pm0.01$	&	F6V	&	3.4768	&	TRUE	&	GJ 216 A	&	Flag B	& $0.34\pm0.02$	\\
43	&	GJ 166 A	&	 63.804843  	&	 -7.672785	&	$5.010\pm0.003$	&	K0V	&	4.1798	&	TRUE	&	GJ 166 A	&	Flag A	 B&	$0.47\pm0.03$\\
44	&	HD 100623 A	&	173.618218	&	-32.826553	&	$9.559\pm0.003$	&	K0V	&	5.7383	&	TRUE	&	GJ 432 A	&	&	$0.06\pm0.01$	\\
45	&	36 Oph C	&	259.052557	&	-26.552693	&	$5.954\pm0.001$	&	K5V	&	5.8898	&	TRUE	&	GJ 664	&	&	$0.84\pm0.04$	\\
46	&	HD 32147	&	75.207377	&	-5.760124	&	$8.844\pm0.002$	&	K3	&	5.8864	&	TRUE	&	GJ 183	&	&	$0.13\pm0.02$	\\
49	&	GJ 66 B	&	24.951421	&	-56.196385	&	$8.189\pm0.002$	&	K2V	&	5.5083	&	TRUE	&	GJ 66 B	&	Flag C	&	$0.52\pm0.03$\\
50	&	chi01 Ori A	&	88.594562	&	20.275727	&	$8.70\pm0.04$	&	G0V	&	4.2416	&	FALSE	&	GJ 222	&	Flag B	& -\\
52	&	ksi UMa Ba	&	169.542832	&	31.525579	&	$8.73\pm0.03$	&	G2V	&	4.6402	&	TRUE	&	GJ 423 Ba	&	Flag B	&	$29.07\pm0.31$\\
53	&	zet Tuc	&	5.041138	&	-64.868014	&	$8.61\pm0.01$	&	F9.5V	&	4.0735	&	TRUE	&	GJ 17	&	Flag B	& $0.09\pm0.01$	\\
57	&	Alpha Centauri A	&	219.85820	&	-60.831234	&	1.34	&	G2V	&	--	&	TRUE	&	GJ 559 A	&	Flag B	 C& $14.93\pm1.64$	\\
58	&	Procyon A	&	114.821329	&	5.218964	&	3.50	&	F5V	&	--	&	TRUE	&	GJ 280	&	Flag B&	$8.20\pm0.21$	\\
59	&	Alpha Centauri B	&	219.85820	&	-60.831228	&	1.34	&	K1V	&	--	&	TRUE	&	GJ 559 B	&	Flag B	 C&	 $14.93\pm1.64$\\
60	&	HD 156384 A	&	259.746199	&	-34.990209	&	6.80	&	K3V	&	--	&	TRUE	&	GJ 667 A	&	Flag B	 C&	$0.80\pm0.04$\\
61	&	HD 156384 B	&	259.746199	&	-34.990209	&	6.80	&	K5V	&	--	&	FALSE	&	GJ 667 B	&	Flag B	 C&	 -\\

\end{longtable}

}
\twocolumn

\end{appendix}
\end{document}